\documentclass[manuscript,screen]{acmart}
\AtBeginDocument{%
  }

\setcopyright{acmlicensed}
\copyrightyear{2025}
\acmYear{2025}
\acmDOI{XXXXXXX.XXXXXXX}
\acmConference[Conference acronym 'XX]{Make sure to enter the correct
  conference title from your rights confirmation email}{June 03--05,
  2018}{Woodstock, NY}
\acmISBN{978-1-4503-XXXX-X/2018/06}

\usepackage{makecell}
\usepackage{multirow}
\usepackage{array}

\usepackage{pifont}
\usepackage{xcolor}
\usepackage{threeparttable}

\begin{document}
\definecolor{red}{RGB}{255,0,0}
%%
%% The "title" command has an optional parameter,
%% allowing the author to define a "short title" to be used in page headers.
\title{Federated Large Language Models: Feasibility, Robustness, Security and Future Directions}

%%
%% The "author" command and its associated commands are used to define
%% the authors and their affiliations.
%% Of note is the shared affiliation of the first two authors, and the
%% "authornote" and "authornotemark" commands
%% used to denote shared contribution to the research.
\author{Wenhao Jiang}
%\authornote{Both authors contributed equally to this research.}
\email{jwh_roy@nudt.edu.cn}
\orcid{0009-0005-1886-3757}
\affiliation{%
  \institution{National University of Defense Technology}
  \country{Changsha, China}
}

\author{Yuchuan Luo}
\email{luoyuchuan09@nudt.edu.cn}
\affiliation{%
  \institution{National University of Defense Technology}
  \country{Changsha, China}
}
\author{Guilin Deng}
\email{dengguilin@nudt.edu.cn}
\affiliation{%
  \institution{National University of Defense Technology}
  \country{Changsha, China}
}
\author{Silong Chen}
\email{chensilong@nudt.edu.cn}

\affiliation{%
  \institution{National University of Defense Technology}
  \country{Changsha, China}
}
\author{Xu Yang}
\email{yangxu16@nudt.edu.cn}
\affiliation{%
  \institution{National University of Defense Technology}
  \country{Changsha, China}
}\author{Shihong Wu}
\email{wushihong@nudt.edu.cn}

\affiliation{%
  \institution{National University of Defense Technology}
  \country{Changsha, China}
}
\author{Xinwen Gao}
\email{gaoxinwen17@nudt.edu.cn}
\affiliation{%
  \institution{National University of Defense Technology}
  \country{Changsha, China}
}

\author{Lin Liu}
\email{liulin16@nudt.edu.cn}
\affiliation{%
  \institution{National University of Defense Technology}
  \country{Changsha, China}
}

\author{Shaojing Fu}
\email{fusj_nudt@163.com}
\affiliation{%
  \institution{National University of Defense Technology}
  \country{Changsha, China}
}
%%
%% By default, the full list of authors will be used in the page
%% headers. Often, this list is too long, and will overlap
%% other information printed in the page headers. This command allows
%% the author to define a more concise list
%% of authors' names for this purpose.
\renewcommand{\shortauthors}{W. Jiang et al.}

%%
%% The abstract is a short summary of the work to be presented in the
%% article.
\begin{abstract}
The integration of Large Language Models (LLMs) and Federated Learning (FL) presents a promising solution for joint training on distributed data while preserving privacy and addressing data silo issues. However, this emerging field, known as Federated Large Language Models (FLLM), faces significant challenges, including communication and computation overheads, heterogeneity,  privacy and security concerns. Current research has primarily focused on the feasibility of FLLM, but future trends are expected to emphasize enhancing system robustness and security. This paper provides a comprehensive review of the latest advancements in FLLM, examining challenges from four critical perspectives: feasibility, robustness, security, and future directions. We present an exhaustive survey of existing studies on FLLM feasibility, introduce methods to enhance robustness in the face of resource, data, and task heterogeneity, and analyze novel risks associated with this integration, including privacy threats and security challenges. We also review the latest developments in defense mechanisms and explore promising future research directions, such as few-shot learning, machine unlearning, and IP protection. This survey highlights the pressing need for further research to enhance system robustness and security while addressing the unique challenges posed by the integration of FL and LLM.
\end{abstract}

%%
%% The code below is generated by the tool at http://dl.acm.org/ccs.cfm.
%% Please copy and paste the code instead of the example below.
%%

%%
%% Keywords. The author(s) should pick words that accurately describe
%% the work being presented. Separate the keywords with commas.

\received{20 February 2007}
\received[revised]{12 March 2009}
\received[accepted]{5 June 2009}

%%
%% This command processes the author and affiliation and title
%% information and builds the first part of the formatted document.
\maketitle
\section{Introduction}
Large language models (LLMs), represented by DeepSeek and ChatGPT, have demonstrated remarkable performance in intelligent question answering, logical reasoning, and natural language processing. This has led to their significant success in various fields, such as biomedicine \cite{li2024open,liu2024fedfms}, legal consulting \cite{zhang2023fedlegal}, and recommendation systems \cite{zhao2024llm,li2024navigating}, and has once again sparked a research boom in artificial intelligence technologies. However, as the scale of these models continues to expand, the demand of raw data for pre-trained models (also known as foundation models \cite{bommasani2021opportunities}, FMs) is also increasing, leading to growing data anxiety. The current trend is that a large amount of high-quality private data is widely distributed among various data holders. Due to privacy concerns, regulation restrictions, or other reasons, these data cannot be publicly shared, resulting in the emergence of data silos \cite{li2022federated}. Villalobos et al. \cite{villalobos2024will} found that high-quality public data will be exhausted before 2026, which will not be able to support the further development of LLMs. Meanwhile, the traditional centralized large-scale collection of private data for training will inevitably violate user privacy.

To address the aforementioned challenges, the integration of LLMs with federated learning (FL) \cite{mcmahan2017communication} emerges as a silver-bullet solution. FL is a distributed machine learning paradigm that enables multiple parties to collaboratively train a single model using their private data while preserving data privacy. This union, known as Federated Large Language Model (FLLM), has recently become a burgeoning research hotspot. Although this powerful combination can overcome data limitations, it also introduces multifaceted challenges, including parameter aggregation for feasibility, heterogeneity for robustness, and privacy and security concerns.

LLMs are a type of FMs, and several recent reviews have explored the prospects of combining FL with FMs \cite{li2024position,li2024synergizing,woisetschlager2024survey,ren2024advances,yu2023federated,zhuang2023foundation}, but only a few have focused on FLLM \cite{hu2024federated,chen2023federated,yao2024federated}. Distinct from these works, we examine FLLM from a novel perspective, focusing on four critical dimensions: Feasibility, Robustness, Security, and Future Directions, with an emphasis on Robustness and Security. Notably, existing surveys, due to the limited pool of early FLLM papers, predominantly cite studies on FL or LLM rather than FLLM itself. In contrast, our work surveys the latest FLLM research, offering readers a more direct and up-to-date understanding of the current research trends. Table ~\ref{tab:diff} provides a comparative overview of our work and some previous surveys.

\begin{table}
\begin{threeparttable}
  \caption{Comparison of our work with the Previous Surveys\tnote{1}}
  \label{tab:diff}
  \begin{tabular}{c|c|c|cccc}
    \toprule
    Year & Work &Topic& Feasibility & Robustness & Security & Future Directions\\
    \midrule
    \multirow{3}{*}{2023} & Chen et al. \cite{chen2023federated} &FLLM& \ding{51} & \ding{55} & \ding{52}\kern-1.2ex\raisebox{1ex}{\rotatebox[origin=c]{125}{\textbf{--}}} & \ding{55}\\
     & Yu et al. \cite{yu2023federated}&FL-FM& \ding{51} & \ding{55} & \ding{55} & \ding{51}\\
     & Zhuang et al. \cite{zhuang2023foundation} &FL-FM& \ding{52}\kern-1.2ex\raisebox{1ex}{\rotatebox[origin=c]{125}{\textbf{--}}} & \ding{55} & \ding{52}\kern-1.2ex\raisebox{1ex}{\rotatebox[origin=c]{125}{\textbf{--}}} & \ding{51}\\
    \hline
    \multirow{7}{*}{2024} & Ren et al. \cite{ren2024advances}&FL-FM& \ding{51} & \ding{55} & \ding{52}\kern-1.2ex\raisebox{1ex}{\rotatebox[origin=c]{125}{\textbf{--}}} & \ding{51}\\
     & Li et al. \cite{li2024synergizing}&FL-FM& \ding{51} & \ding{51} & \ding{52}\kern-1.2ex\raisebox{1ex}{\rotatebox[origin=c]{125}{\textbf{--}}} & \ding{51}\\
     & Woisetschlager et al. \cite{woisetschlager2024survey}&FL-FM& \ding{51} & \ding{55} & \ding{55} & \ding{52}\kern-1.2ex\raisebox{1ex}{\rotatebox[origin=c]{125}{\textbf{--}}} \\
     & Li et al. \cite{li2024position}&FL-FM& \ding{51} & \ding{55} & \ding{52}\kern-1.2ex\raisebox{1ex}{\rotatebox[origin=c]{125}{\textbf{--}}} & \ding{51}\\
     & Yao et al. \cite{yao2024federated}&FLLM& \ding{51} & \ding{51} & \ding{52}\kern-1.2ex\raisebox{1ex}{\rotatebox[origin=c]{125}{\textbf{--}}} & \ding{51}\\
     & Hu et al. \cite{hu2024federated}&FLLM&\ding{51} & \ding{55} & \ding{52}\kern-1.2ex\raisebox{1ex}{\rotatebox[origin=c]{125}{\textbf{--}}}& \ding{52}\kern-1.2ex\raisebox{1ex}{\rotatebox[origin=c]{125}{\textbf{--}}} \\
     \hline
     2025 & \textcolor{red}{Our work} &FLLM& \ding{51} & \ding{51} & \ding{51} & \ding{51} \\
  \bottomrule
\end{tabular}
\begin{tablenotes}
\item[1] \ding{51} represents high correlation, \ding{52}\kern-1.2ex\raisebox{1ex}{\rotatebox[origin=c]{125}{\textbf{--}}} represents few references to the FLLM paper, and \ding{55} represents no relevant content.
\end{tablenotes}
\end{threeparttable}
\end{table}

To address the feasibility of FLLM, it is essential to explore efficient implementation methods. Training LLMs involves updating parameters at the trillion scale, which generates substantial communication and computational overheads that are unacceptable for resource-constrained federated participants. To tackle this issue, Federated Parameter-Efficient Fine-Tuning (FedPEFT) \cite{malaviya2023reducing} has emerged as a promising solution. Based on extensive research, we categorize the fine-tuning methods for FLLM into four types: full-parameter fine-tuning, parameter-efficient fine-tuning, prompt tuning, and other specialized fine-tuning methods. Currently, numerous studies have addressed the feasibility of FLLM from an academic perspective, demonstrating the potential for fine-tuning LLMs on client devices.

Robustness corresponds to the heterogeneity of FLLM. Unlike centralized training, where data is aggregated in a single location, FLLM is trained on discretely distributed data across various clients, each equipped with distinct hardware and software environments. This setup readily gives rise to heterogeneity issues, which we categorize into three types: resource heterogeneity, data heterogeneity, and task heterogeneity. (\romannumeral1) Resource heterogeneity refers to the differences in the computational and storage resources available to clients, which affects the training process and aggregation strategies. Unlike traditional FL, the large scale of parameters trained during FLLM means that resource-constrained clients may not be able to efficiently complete the fine-tuning tasks, resulting in slow training processes and even incomplete training. (\romannumeral2) Data heterogeneity refers to the unequal distribution of data across clients, that is, the data are non-independent and identically distributed (Non-i.i.d.) \cite{chen2024feddat}. Different clients may have different update directions, leading to drift in the global model update. Due to the complex structure and strong memory of the model resulting from its parameters, FLLM is more affected by data heterogeneity compared to traditional FL, leading to greater instability. (\romannumeral3) Task heterogeneity indicates that different clients may have different types of tasks \cite{chen2024feddat,yang2024dual,ping2024fl,bai2024federated,cho2024heterogeneous}. Given the strong semantic understanding capabilities of LLMs, they can be applied to a wide range of tasks, such as question-answering, classification, text generation, and translation. This diversity in tasks across clients leads to significantly different convergence directions for the models, creating obstacles to the robustness of FLLM. Currently, several studies have focused on addressing the heterogeneity issues in FLLM to enhance their robustness. However, more in-depth research is still needed to tackle the challenges posed by data and task heterogeneity.

\begin{figure}[htbp]
  \centering
  \includegraphics[width=0.9\linewidth]{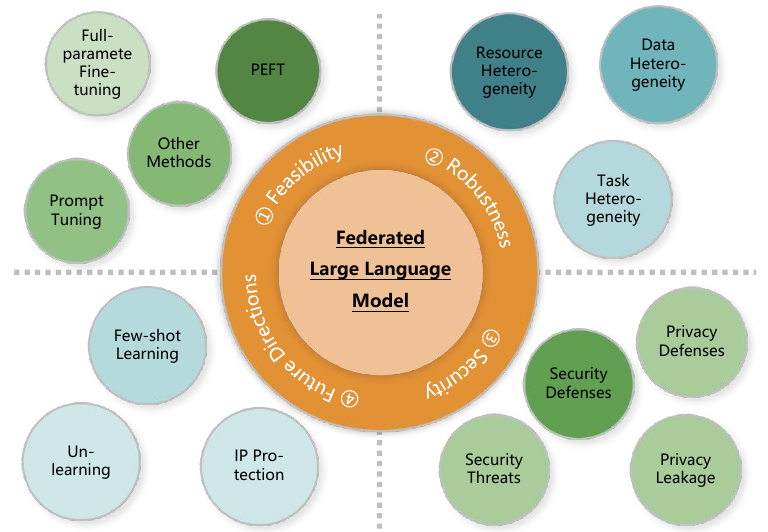}
  \caption{The framework of this survey. Darker circles indicate a higher number of related studies.}
  \label{kuangjia}
\end{figure}

Security corresponds to the privacy and security of FLLM. Integrating FL into LLMs may simultaneously introduce privacy threats from both LLMs and FL, such as data reconstruction and membership inference\cite{chen2023federated,han2024fedsecurity}. Security challenges mainly refer to the threats posed by malicious attackers or curious participants exploiting vulnerabilities to impair system performance. In the traditional FL, these primarily include model poisoning attacks\cite{Ma2023Poisoning}, data poisoning attacks\cite{Fabio2013Evasion}, backdoor attacks\cite{Zuo2024Byzantine}, and adversarial attacks\cite{Zhou2024Fair}. In FLLM, textual data is characterized by its discrete nature, where each word or character corresponds to a discrete index. For gradient leakage attacks, this discreteness makes it more challenging to directly reconstruct the original text from the gradient information. For poisoning attacks, the complexity of the model and the discreteness of the data require more sophisticated strategies for the attacker to achieve effective poisoning. For inference attacks, attackers need to infer certain features of the text by analyzing the model's outputs or the activations of intermediate layers. These attacks may either affect the model convergence speed or even prevent convergence, degrade model performance, or cause incorrect inference results on specific data, ultimately reducing the credibility of the model. 

The future direction primarily focuses on the issues that remain unexplored by FLLM. For example, making full use of scarce training data, protecting user rights, and protecting the intellectual property (IP) of models are additional requirements for FLLM to achieve long-term development. The significance of studying few-shot scenarios is substantial, as acquiring large amounts of labeled data is both time-consuming and costly. In training FLLM, insufficient labeled data can prevent the model from fully learning and understanding the patterns and features within the data, thereby affecting the model's training effectiveness and generalization ability. Moreover, the issue of catastrophic forgetting, which may arise as FLLM meets the need for clients to delete private training data, poses a significant barrier to its development. Furthermore, in terms of intellectual property protection, ensuring that the model is not illegally copied, redistributed, or misused has become a crucial challenge for FLLM.

In this survey, we focus primarily on recent papers concerning FLLM from four aspects: feasibility, robustness, security, and future directions. We present the framework of our survey in Figure ~\ref{kuangjia}, where the darker circles indicate a larger number of related studies. Unlike previous reviews, we place particular emphasis on the novel issues of heterogeneity and privacy and security arising from the integration of FL and LLMs, as well as the corresponding solutions, and have collected the latest FLLM-related papers. Our goal is to provide valuable references for researchers and practitioners interested in this interdisciplinary field. We systematically review the latest methods, highlight their contributions, and discuss how they address the inherent challenges of combining FL with LLM. Finally, we emphasize potential research directions that may emerge in the near future and summarize the entire text.

\section{Feasibility of FLLM}
FLLM essentially involves training LLMs using the methodology of FL. The conventional process of training an LLM comprises two primary stages: pre-training and fine-tuning. Currently, the number of parameters in LLMs has reached hundreds of billions and continues to increase. Companies such as Google, OpenAI, and Huawei have all proposed the development of models with trillions of parameters. For the majority of clients participating in FLLM, they lack the abundant resources of a server, making it impractical to jointly train LLMs from scratch. Therefore, FLLM predominantly focuses on the fine-tuning phase. Under the premise of preserving privacy, fine-tuning FLLM with vertical domain data represents the optimal approach for fine-tuning on private data for downstream tasks. However, in terms of computational costs, the substantial computational demands of backpropagation make it challenging for clients to fine-tune LLMs. Even with GPU clusters, training such models remains a formidable challenge. Regarding communication costs, since FLLM necessitates the sharing of model gradients or aggregated model parameters, the limited network bandwidth of clients and the extensive transmission of parameters can lead to communication bottlenecks between clients and the server.

\begin{figure}[htbp]
  \centering
  \includegraphics[width=\linewidth]{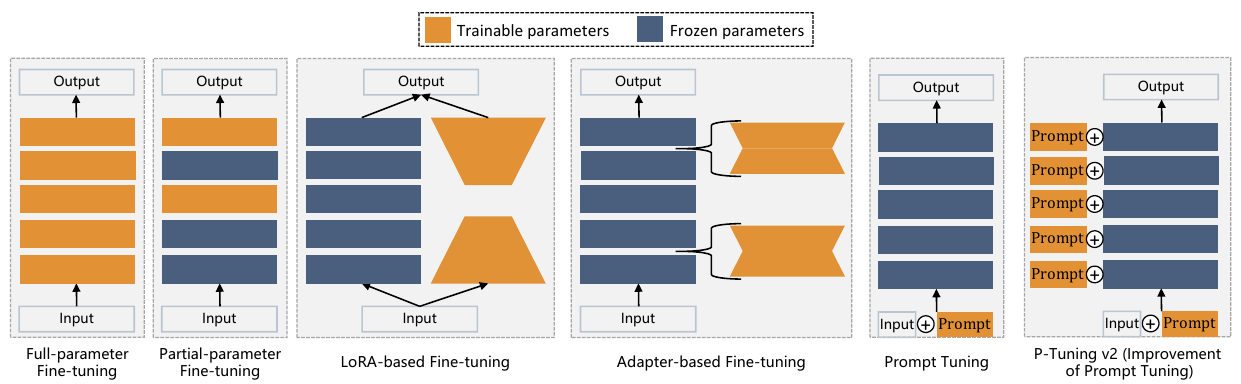}
  \caption{The fine-tuning process of full-parameter fine-tuning, PEFT (including partial-parameter fine-tuning, LoRA, Adapter) and prompt tuning. The orange box denotes the trainable parameters.}
  \label{tuning}
\end{figure}

To overcome the challenges posed by computational and communication costs and to realize the feasibility of FLLM, a series of Federated Parameter-Efficient Fine-Tuning (FedPEFT) methods have continuously emerged \cite{malaviya2023reducing}. In this section, we categorize the training methods according to the modifications made during the training process of FLLM into full-parameter fine-tuning, parameter-efficient fine-tuning, prompt tuning, and other special techniques, and introduce each method respectively, as shown in Table ~\ref{tab:Feasibility}. To clearly delineate the distinctions among various fine-tuning methods, we have indicated the trainable parameters in orange in Figure ~\ref{tuning}.

\begin{table}[htbp]
\renewcommand{\arraystretch}{1.2}
\caption{Overview of feasibility in FLLM.}
\begin{center}
\addtolength{\tabcolsep}{0pt}
\begin{tabular}{c|c|c}
\hline

\textbf{Type}& \textbf{Describe} & \textbf{Approaches} \\
\hline
\makecell{Full-parameter\\Fine-tuning}& \makecell{Fine tune all of the parameters} &\makecell{Fedlegal \cite{zhang2023fedlegal}, FedRDMA \cite{zhang2024fedrdma}}\\
\hline

\makecell{Parameter-Efficient\\Fine-tuning (PEFT)}& \makecell{Fine tune part of the parameters,\\or freeze parameters of LLM and\\ insert additional trainable modules.} &\makecell{FedCyBGD \cite{wang2024saveallenablingparameter}, BitFit \cite{zaken2022bitfitsimpleparameterefficientfinetuning},\\ FeS \cite{Mengwei2023Federated}, FedSelect \cite{Shamsian2024FedSelect},\\ AdaFL \cite{cai2023efficient}, FlexLoRA \cite{bai2024federated}, \\HETLORA \cite{cho2024heterogeneous}, FwdLLM \cite{xu2024fwdllm},\\ LP-FL \cite{jiang2024low}, DP-DyLoRA \cite{xu2024dp}}\\

\hline
Prompt tuning&\makecell{Fine-tunes prompts without altering \\the parameters of LLMs.} &\makecell{Promptfl \cite{guo2023promptfl}, FedTPG \cite{qiu2023text},\\ TCFL \cite{zhao2023inclusive}} \\
\hline
Other methods & \makecell{Model compression, split learning,\\ and zeroth-order optimization.} &\makecell{\cite{10097124,yu2023bridging,wu2024fedbiot}, \cite{lin2024splitlora,zheng2024safely}, \cite{qin2023federated,ling2024convergence,lu2023zoopfl}} \\

\hline

\end{tabular}
\label{tab:Feasibility}
\end{center}
\end{table}

\subsection{Full-parameter Fine-tuning}

Full-parameter fine-tuning refers to the process of continuing to train the parameters of a pre-trained LLM on a private dataset, which involves updating all parameters. During fine-tuning, the model architecture remains unchanged, and only the existing parameters are optimized to adapt to specific downstream tasks.

Full-parameter fine-tuning is the most straightforward method of parameter fine-tuning, where all model parameters are updated during the fine-tuning process. Zhang et al.\cite{zhang2023fedlegal} achieved the first FLLM in the legal domain by federated fine-tuning the RoBERTa-WWM pre-trained model released by HuggingFace on private information. However, the communication and computational costs associated with this method are highly prominent issues. To address this, Zhang et al.\cite{zhang2024fedrdma} integrated RDMA technology into the FL framework to improve communication efficiency and robustness, thereby realizing the FedRDMA framework for FLLM fine-tuning based on an industrial FL framework. Compared with traditional TCP/IP-based FL systems, FedRDMA improves communication efficiency by 3.8 times. Nevertheless, this framework merely enhances communication efficiency through RDMA technology without resolving the significant computational costs, storage demands, and communication overheads associated with full-parameter fine-tuning. Wang et al.\cite{wang2024saveallenablingparameter} proposed a full-parameter fine-tuning method called FedCyBGD to tackle these issues. This method uses cyclic block gradient descent to divide the model into multiple blocks, with each client responsible for updating one or several specific blocks cyclically, rather than the entire model, thereby reducing computational and storage costs. In addition, a model compression scheme was designed to lower communication costs, enabling full-parameter fine-tuning with lower resource consumption. However, this approach lacks sufficient convergence analysis and is limited by the small number of clients considered in its experiments. Due to the inability of full-parameter fine-tuning to fundamentally solve the problems of high computational costs, storage demands, and significant communication overheads, independent research on this method is relatively rare, and it is not the preferred approach for FLLM. Instead, it is often used as a reference and comparison for other methods \cite{zaken2022bitfitsimpleparameterefficientfinetuning}. 

The advantages and disadvantages of full-parameter fine-tuning are both quite evident. This fine-tuning approach is highly adaptable, allowing for fine-grained adjustments to the model's internal representations to better fit downstream tasks. When sufficient resources are available, Full-parameter fine-tuning typically yields the best performance, making it widely applicable across various model architectures and task types. However, it requires updating and transmitting a large number of parameters during each communication round, demanding high hardware requirements for the devices. Moreover, with limited data, overfitting is likely to occur, which in turn degrades the model's generalization ability.

\subsection{Parameter-Efficient Fine-Tuning}

Federated Parameter-Efficient Fine-Tuning (FedPEFT) \cite{malaviya2023reducing} is a series of methods aimed at reducing the computational and communication overhead of federated fine-tuning. We categorize these methods into three types based on the part being updated: partial-parameter fine-tuning, LoRA-based fine-tuning and Adapter-based fine-tuning.

Partial-parameter fine-tuning methods improve the full-parameter fine-tuning methods, reducing computational, storage, and communication costs by decreasing the number of trained parameters. Zaken et al. \cite{zaken2022bitfitsimpleparameterefficientfinetuning} proposed BitFit, a sparse fine-tuning method that updates only the bias parameters. This method achieves comparable or even better performance than fine-tuning the entire model with significantly fewer parameters on small-scale to medium-sized datasets, as full-parameter fine-tuning is prone to overfitting. However, the theoretical explanation for BitFit is insufficient, and whether the selection of bias terms is optimal remains questionable. Sun et al. \cite{sun2024conqueringcommunicationconstraintsenable} also focused on reducing overhead by updating only the bias part of the globally shared model, achieving good results. Cai et al. \cite{Mengwei2023Federated} extended this idea by proposing the FeS framework, which updates only the bias parameters of the intermediate layers and freezes the weights of the lower layers, resulting in improved training efficiency while maintaining good model performance. Rishub Tamirisa et al. \cite{Shamsian2024FedSelect} were inspired by the Lottery Ticket Hypothesis to introduce a method for finding the optimal parameters for local fine-tuning while freezing the remaining parameters. They use gradient information to identify the parameters that change the least during training, which are considered suitable for freezing and carrying shared knowledge, while parameters with significant changes are considered the best choices for local fine-tuning.

\begin{figure}[htbp]
  \centering
  \includegraphics[width=\linewidth]{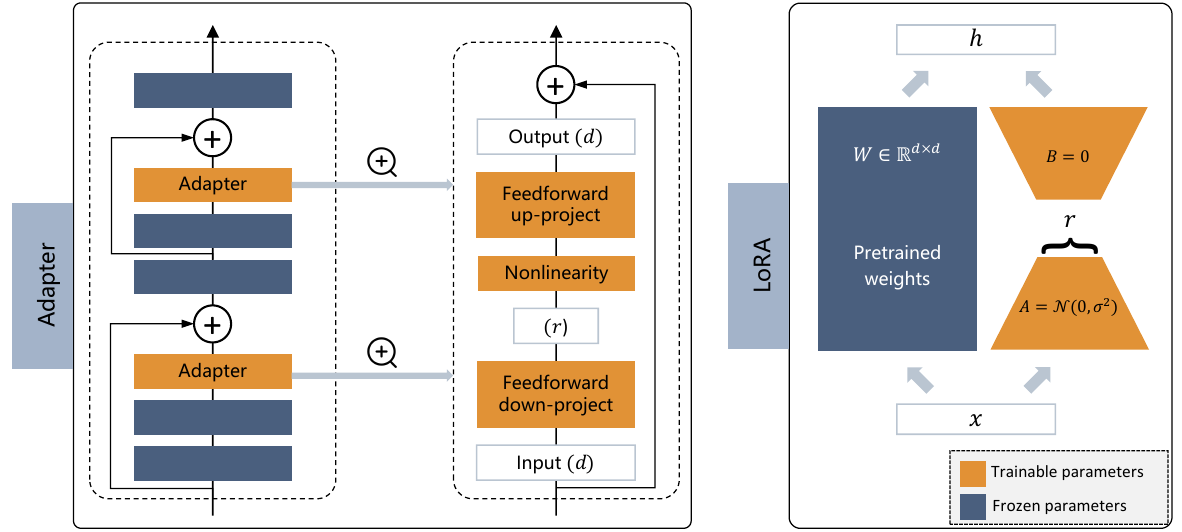}
  \caption{The process of Adapter-based Fine-tuning and LoRA-based Fine-tuning.}
  \label{loraanda}
\end{figure}

Adapter \cite{houlsby2019parameter} and LoRA \cite{hu2021lora} (a.k.a. Reparameterization) freeze the entire pre-trained LLM and introduce some trainable small modules at different positions of the model structure to achieve model fine-tuning for downstream tasks. These small trainable modules usually have low-rank characteristics, which can significantly reduce the number of training parameters and thus lower the computational, storage, and communication costs. Adapter is inserted into the model in a “serial” manner, while LoRA is inserted in a “parallel” manner. Therefore, it is generally believed that adapter will cause additional inference latency, while LoRA will not introduce new inference latency. Adapter-based fine-tuning inserts small adapter modules into each layer of the pre-trained model and fine-tunes only the parameters of these adapter modules to adapt to downstream tasks, as shown on the left side of Figure ~\ref{loraanda}. LoRA (Low-Rank Adaptation) introduces low-rank decomposition matrices into specific layers of LLMs, as shown on the right side of Figure ~\ref{loraanda}. For a weight matrix $W$, LoRA represents its update as $\Delta W = BA$, where $B$ and $A$ are low-rank matrices, and $r \ll d$ (where $d$ is the dimension of the original matrix, and $r$ is the dimension of the low-rank matrix). During training, the original weight matrix $W$ is frozen, and only the low-rank matrices $B$ and $A$ are trained. Thus, during forward propagation, the model's output can be expressed as $h = Wx + BAx$. 

In FLLM with adapters, after the client training is completed, the adapter parameters are sent to the server for aggregation, and then the aggregated adapter parameters are broadcast back to the clients for the next round of training. Cai et al. \cite{cai2023efficient} proposed a FLLM scheme with dynamically configured adapters based on the adapter configuration method in \cite{pfeiffer2020adapterfusion}, and provided a search method for the optimal adapter configuration, supplemented by cache activation technology to achieve efficient federated fine-tuning of pre-trained LLMs. Subsequently, they proposed the FedAdapter \cite{cai2022fedadapter} scheme, which inserts adapters into different layers of the pre-trained model for training, adopts a dynamic configuration mechanism to adjust the depth and width of the adapters according to the training progress, and further optimizes the computational efficiency of the clients through activation caching technology. However, the dynamic configuration process of adapters in FedAdapter requires additional computational and communication overhead, and the configuration space of adapters (depth and width) may not cover all optimization solutions. FedOA \cite{yang2025federated} enhances the model's generalization ability in out-of-distribution scenarios by combining adapters with feature distance regularization techniques and the global model's invariant feature learning capability, effectively addressing the challenges posed by large-scale parameters and data heterogeneity in FLLM. However, its convergence speed and stability in practical applications may be affected by the client data distribution and communication frequency, and it is sensitive to the selection of hyperparameters (such as regularization strength $\lambda$). Jia et al. \cite{jia2024efficientmodelheterogeneityfederatedlearning} proposed the HeteroTune scheme, combining multi-branch cross-model aggregators to achieve efficient knowledge aggregation of models of different sizes on heterogeneous devices.

For FLLM with LoRA, different frameworks have different implementation processes. For example, in the FedLoRA framework \cite{yi2023fedlora}, clients upload their trained $A$ and $B$ matrices to the server, which aggregates these matrices to generate global $A$ and $B$ matrices and broadcasts them back to the clients. Clients then use the global $A$ and $B$ matrices for further optimization in the next round of training. Using this LoRA fine-tuning method, Jiang et al. \cite{jiang2024low} implemented federated fine-tuning of pre-trained LLMs and found through experimental comparisons that this method can achieve performance similar to or even better than full-parameter fine-tuning. Bai et al. \cite{bai2024federated} designed the FlexLoRA method, which allows clients to dynamically adjust the rank of LoRA matrices based on local resources to change the number of trainable parameters. After aggregating the global LoRA, the server uses singular value decomposition to redistribute the parameters. Cho et al. \cite{cho2024heterogeneous}, based on LoRA fine-tuning, employed techniques such as rank self-pruning and sparse weighted aggregation to further improve the convergence speed of the global model. Fang et al. \cite{fang2024automated} designed methods for identifying trainable weight importance and fast search algorithms to quickly search for the optimal low-rank adaptive matrices locally and further reduced storage requirements using quantization techniques. To further reduce memory overhead, researchers have integrated techniques such as forward differentiation with LoRA fine-tuning \cite{xu2024fwdllm, panchal2024thinking, xu2024dp}.

Adapter and LoRA methods are both representative techniques of PEFT and are suitable for fine-tuning LLMs in FL. Adapter inserts lightweight modules and is suitable for a variety of tasks (such as NLP, CV) and heterogeneous data distribution scenarios, with performance close to full model fine-tuning, but with higher communication overhead; LoRA, on the other hand, decomposes weight matrices into low-rank matrices and is particularly suitable for NLP tasks and homogeneous data distribution scenarios, with low communication overhead, but may have limited expressive power in complex tasks. Overall, adapter is more suitable for diverse tasks and heterogeneous data, while LoRA performs better in NLP tasks and scenarios with low communication demands.

\subsection{Prompt tuning}

Prompt tuning can fine-tune continuous prompts(a.k.a. soft prompts) without altering the parameters of the LLM \cite{lester2021power}, and is an emerging promising method for reducing costs and protecting privacy. The embedding layer, which is the initial component of the model, maps discrete words or tokens into a continuous vector space. These embedding vectors capture the semantics and contextual information of the words, forming the basis for the model's subsequent processing. The feasibility of prompt tuning is attributed to the fact that pre-trained language models (such as GPT, BERT) have acquired extensive linguistic knowledge and universal semantic representations through unsupervised learning on large-scale corpora. This universality provides the foundation for prompt tuning. Essentially, a prompt serves as a guiding signal, which directs the model to generate the desired output by incorporating specific text segments or structures into the input text. The core of prompt tuning lies in utilizing the model's embedding layer to implement trainable prompts, which are adjusted through optimization methods such as gradient descent to better adapt to specific tasks. During the process of prompt tuning, a set of prompt embedding vectors generated via the model's embedding layer is first initialized. Subsequently, these prompt embedding vectors are combined with task-specific input data to form the complete input. By optimizing the prompt embedding vectors using labeled data, the model can better adapt to specific tasks. In the inference stage, the optimized prompt embedding vectors are employed to generate the output for the task, thereby enhancing the model's performance on specific tasks. Since the number of parameters in the prompt embedding vectors is relatively small, updating these parameters is much more efficient than updating the entire model's parameters, making prompt tuning feasible on resource-constrained devices. As for FLLM, the classical approach of prompt tuning involves each client generating specific prompts for its local data and fine-tuning these prompts. Subsequently, the updated prompts are transmitted to the server for aggregation. The server then broadcasts the aggregated prompts back to the clients for the next round of training, as proposed in the referenced paper \cite{guo2023promptfl}.

\begin{figure}[htbp]
  \centering
  \includegraphics[width=\linewidth]{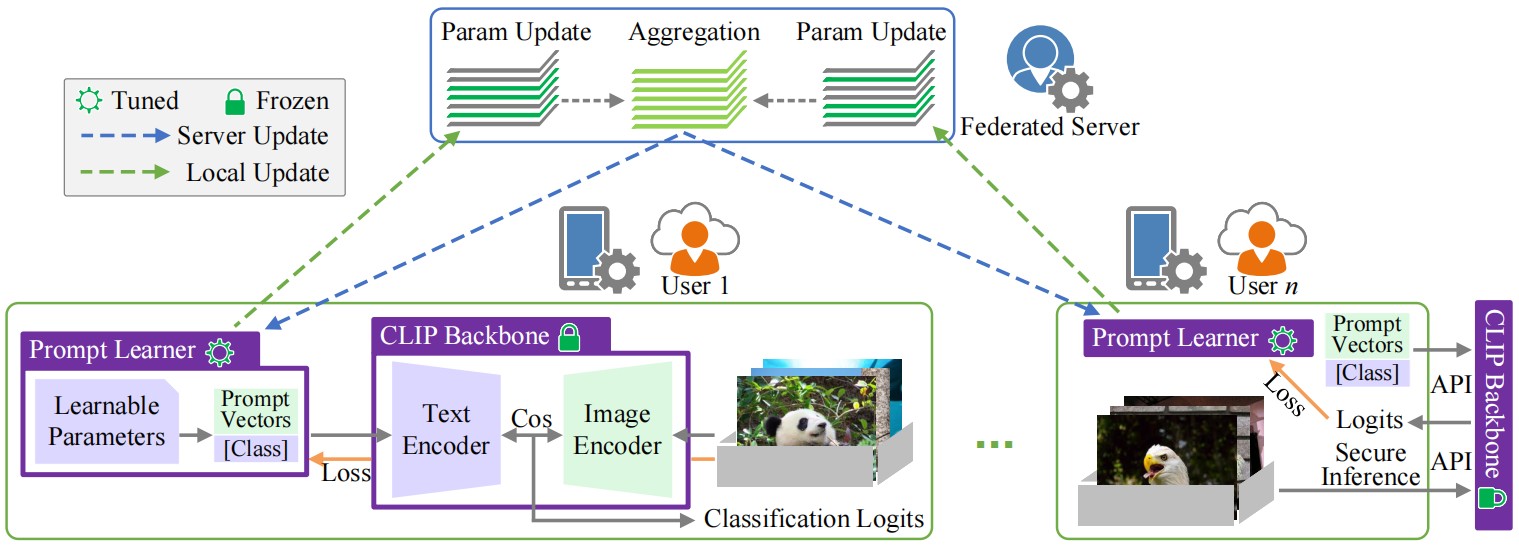}
  \caption{The framework and workflow of PROMPTFL from \cite{guo2023promptfl}. Clients locally train a prompt learner with a small number of parameters, without altering the original LLM, and the server aggregates only the updates from the prompt learners.}
  \label{prompt}
\end{figure}

The effectiveness of training prompts stems from the fact that LLMs are already quite intelligent. Pre-trained models (such as GPT, BERT, etc.) have already learned a vast amount of linguistic knowledge, but they require a clear "guidance" to complete specific tasks. The prompt serves as this "guidance." Training prompts only adjusts the prompt embeddings, not the parameters of the entire model, much like a teacher instructing a student on how to understand a question, rather than reteaching the student all the knowledge. Guo et al. \cite{guo2023promptfl} generated a prompt learner on the client side, transforming federated model training into federated prompt training. Taking the pre-trained LLM CLIP \cite{radford2021learning} as an example, clients train soft prompts with a small amount of local data, send the updates of the prompt learner to the server for aggregation, and then update the local prompt learner based on the feedback from the server, as shown in Figure ~\ref{prompt}. This method of prompt tuning only requires training a small-scale prompt learner, without the need to fine-tune the LLM itself, thereby significantly reducing computational and communication costs. Qiu et al. \cite{qiu2023text} proposed the FedTPG scheme, which involves learning a unified prompt generation network (Prompt Generator) globally to convert task-related text inputs into context-aware prompt vectors. This approach transforms the training of LLMs into the training of the prompt generation network, not only enhancing the model's adaptability but also maintaining low communication costs, making it suitable for FLLM scenarios. Zhao et al. \cite{zhao2023inclusive} proposed the TCFL scheme. This dual-prompt FL integrates visual and textual modalities to overcome the limitations of single-modality prompt tuning in FL, thereby improving data representation among nodes.

In addition, P-Tuning v2 \cite{2110-07602} is an improved deep prompt tuning method that aims to address the limitations of traditional prompt tuning across models of varying sizes and tasks. By introducing trainable prompt embeddings at each layer of the model, it significantly enhances model performance, enabling it to match fine-tuning in a variety of natural language understanding (NLU) tasks.

The advantages of prompt tuning lie in its efficiency, privacy protection, as well as adaptability and flexibility. Given that the number of parameters in prompt embedding vectors is relatively small, updating these parameters is much more efficient than updating the entire model's parameters. This makes prompt tuning feasible on resource-constrained devices. In FLLM, the updates of prompt embedding vectors can protect the privacy of user data, as clients only need to share the updates of prompt embedding vectors, rather than the original data. Moreover, prompt embedding vectors are continuous in the embedding space and can be freely adjusted, without being restricted by natural language vocabulary. This flexibility enables prompts to better capture the complexity of tasks. Through careful design and optimization of prompts, prompt tuning can enable large language models to perform well in a variety of application scenarios.

\subsection{Other Methods}

In addition to parameter fine-tuning and prompt tuning, several other techniques have also contributed to the feasibility of FLLM. For instance, the introduction of model compression, split learning, and zeroth-order optimization has reduced the computational and communication overhead of FLLM, further facilitating its practical implementation.

Model compression minimizes the size of LLMs without compromising performance, thereby reducing the number of parameters that need to be trained \cite{10097124, yu2023bridging, wu2024fedbiot}. This approach effectively reduces computational, memory, and communication costs. Researchers from Google, including Yang et al. \cite{10097124}, proposed the Online Model Compression (OMC) technique for lightweight operations, which compresses model parameters. Parameters are only decompressed and released into memory when they are involved in computations, and the model is shared and stored in a compressed format, thus reducing storage and communication costs. Model parameter quantization techniques reduce the storage and computational resources required by models by lowering the representation precision of model parameters, thereby reducing resource costs\cite{10097124, fang2024automated, zhu2024promoting}. Yang et al. \cite{10097124} studied the impact of parameter quantization on model performance from three different dimensions: full parameter quantization, weight-only quantization, and partial variable quantization, demonstrating the feasibility of quantization techniques.

Split learning approaches the problem from the perspective of model architecture by partitioning the LLM into several sub-models of varying sizes, which are then distributed across the server and clients. This ensures that the primary computational load during training remains on the server. This method significantly reduces the resource demands on client devices in FLLM and has been applied in FLLM \cite{lin2024splitlora, zheng2024safely}.

Zeroth-order optimization techniques \cite{qin2023federated, ling2024convergence, lu2023zoopfl} consider the training process, achieving the feasibility of FLLM by adjusting the backpropagation process during training, optimizing model parameters without directly computing gradients, greatly reducing time costs. Noting that the resource demands of neural network training are significantly higher than those of inference, Xu et al. \cite{xu2024fwdllm} proposed an efficient federated fine-tuning of LLMs based on perturbation inference without the backpropagation process. This method uses forward inference to determine the correctness of perturbations and combines specially designed perturbation discrimination methods to quickly eliminate perturbations that are almost orthogonal to the true gradient, achieving efficient training. Despite the efficient perturbation discrimination method, forward inference still needs to be executed multiple times to obtain an unbiased estimate of the true gradient. Panchal et al. \cite{panchal2024thinking} proposed estimating gradients by computing Jacobian-Vector Products based on random perturbations of weights during the forward pass. This allows gradient estimation to be completed with only two forward inferences, further improving training efficiency. 

The feasibility of FLLM remains a recent research hotspot. At its core, the issue is to address the significant overhead associated with FLLM. The solutions we discussed have basically achieved the integration of FL and LLM in academia. Currently, FedPEFT methods are evolving towards lower computational, storage, and communication costs, enhancing the efficiency of FLLM on the basis of achieving feasibility.

\section{Robustness of FLLM }

Traditional FL already faces challenges of resource heterogeneity\cite{li2021align} and data heterogeneity\cite{mendieta2022local} due to the distributed nature of training data across numerous clients with varying label distributions, sample sizes, and diverse device and network environments. Owing to the substantial parameter scale of LLMs, FLLM significantly exacerbates the issue of resource heterogeneity. Edge nodes with limited resources may not be able to participate in training normally due to computational and communication challenges. LLMs are highly complex and more sensitive to the quality and quantity of data, thereby exacerbating the impact of data heterogeneity. Moreover, FLLM introduces a novel issue: task heterogeneity. \cite{bai2024federated}. This arises because LLMs possess strong semantic understanding capabilities and can be applied to a wide variety of tasks, leading to different clients having distinct tasks such as question-answering, classification, text generation, and translation. The heterogeneous tasks on clients result in significantly different model convergence directions, which is markedly different from traditional FL. Heterogeneity issues are significant barriers to the practical deployment of FLLM, profoundly affecting their robustness. We have conducted a survey of recent works on resource heterogeneity, data heterogeneity, and task heterogeneity in FLLM, as shown in Table ~\ref{tab:Applicability}.

\begin{table*}[htbp]
\renewcommand{\arraystretch}{1.2}
\caption{Overview of robustness in FLLM.}
\begin{center}
\addtolength{\tabcolsep}{0pt}
\begin{tabular}{c|c|c}
\hline

\textbf{Type}& \textbf{Describe} & \textbf{Approaches} \\
\hline
Resource Heterogeneity& \makecell{Diverse computational resources impact\\ the aggregation efficiency of FLLM.} &\makecell{FedLoRA \cite{yi2023fedlora}, FlexLoRA \cite{bai2024federated},\\ Fed-piLot \cite{zhang2024fed}, HETLORA \cite{cho2024heterogeneous},\\ FedRA \cite{su2025fedra}, FedSpaLLM \cite{bai2024fedspallm}}\\
\hline
Data Heterogeneity&\makecell{Unbalanced data distribution\\ leads to global model drift.}  &\makecell{FDLoRA \cite{qi2024fdlora}, FedPipe \cite{fang2024automated},\\ PFIT \cite{jiang2024personalized}, SPRY \cite{panchal2024thinking},\\ FwdLLM \cite{mei2024fedmoe}}\\
\hline
Task Heterogeneity&\makecell{The demand for multitasking makes it\\ difficult for FLLM to converge stably.} &\makecell{M2FEDSA \cite{zhangenhancing}, FedDPA \cite{yang2024dual},\\ FedBone \cite{chen2024fedbone}, FedDAT \cite{chen2024feddat},\\ FL-TAC \cite{ping2024fl}} \\
\hline

\end{tabular}
\label{tab:Applicability}
\end{center}
\end{table*}

\subsection{Resource Heterogeneity}
In reality, FLLM systems are highly likely to be composed of devices with heterogeneous resources, featuring diverse software and hardware environments. This leads to differences among clients in terms of computing platforms, computational capabilities, storage space, communication bandwidth, and training efficiency, thereby affecting the training process and aggregation strategies \cite{qiu2024federated}. FLLM is also subject to the "bucket effect," where efficiency is determined by the least capable participant, resulting in the waste of resources from other participants. Given the widespread existence of resource heterogeneity, existing research on the heterogeneity of FLLM mainly focuses on this issue \cite{bai2024federated, zhang2024fed, su2025fedra, cho2024heterogeneous, thapa2022splitfed, tian2022fedbert, shin2023fedsplitx, singh2019detailed, xu2023federated, bai2024fedspallm, su2024federated}. In these studies, most methods are based on LoRA and adapter to address resource heterogeneity.

The PEFT methods mentioned above not only contribute to the feasibility of FLLM, but also enhance their robustness. The FedLoRA framework \cite{yi2023fedlora} mentioned previously achieves parameter-efficient fine-tuning of large heterogeneous models by inserting a small LoRA on each client and aggregates LoRA on the server side to enable knowledge sharing among clients. Zhang et al. \cite{zhang2024fed} proposed the Fed-piLot scheme, which, based on the observation that training different LoRA layers results in different memory consumption and that different layers contribute differently to model performance, formulated the allocation of LoRA as a knapsack optimization problem. They designed a value function based on local-global information gain score (IG-Score) to optimize the allocation of LoRA under client memory constraints. Cho et al. \cite{cho2024heterogeneous} first discussed that in resource-heterogeneous scenarios, the redistribution aggregation method using LoRA faces problems of overfitting and slow convergence, and thus proposed the HETLORA scheme. By applying local rank self-pruning and sparse weighted aggregation on the server, it combines the advantages of high- and low-rank LoRAs, achieving better convergence speed and final performance compared to homogeneous LoRA. Su et al. \cite{su2025fedra} proposed the FedRA scheme, which randomly generates an allocation matrix in each communication round to determine which layers of the model each client is responsible for updating. Resource-constrained clients only need to handle a small number of layers assigned to them and fine-tune through adapters, thereby reducing computational overhead. The server aggregates the adapter updates from clients to the corresponding layers of the global model according to the allocation matrix. This strategy not only fully utilizes the computational resources of different clients but also supports extremely heterogeneous scenarios where no client can fine-tune the entire model. 

\begin{figure}[h]
  \centering
  \includegraphics[width=0.7\linewidth]{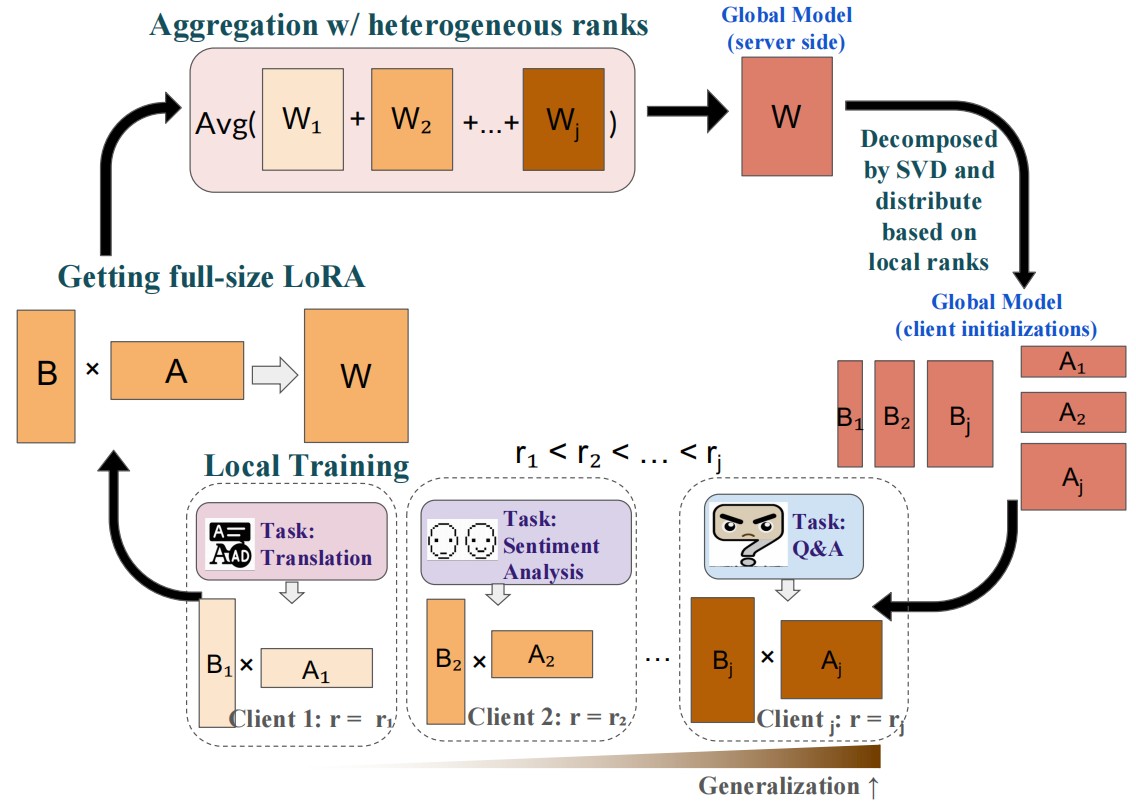}
  \caption{FlexLoRA workflow from \cite{bai2024federated}. The server aggregates the full-size LoRA after multiplication, not the individual matrices A and B. The global full-size LoRA is then decomposed into smaller matrices of varying ranks via SVD and allocated sequentially based on client resources.}
  \label{flexlora}
\end{figure}

Furthermore, Bai et al. \cite{bai2024federated} proposed the FlexLoRA scheme, which allows clients to dynamically adjust local LoRA ranks based on their own resources, as shown in Figure ~\ref{flexlora}. Unlike previous methods where the server aggregated matrices A and B as $B_g=(\sum_{i=1}^mn^iB^i_l)/(\sum_{i=1}^mn^i)$, $A_g=(\sum_{i=1}^mn^iA^i_l)/(\sum_{i=1}^mn^i)$, where $m$ is the number of FL clients, $n_i$ is the size of the i-th client’s local training dataset, $B_g$, $A_g$ are the global LoRA decomposed matrices, and $B^i_l$, $A^i_l$ are the local LoRA decomposed matrices of i-th client, FlexLoRA aggregates the full-size LoRA $W_g=(\sum_{i=1}^mn^iW_l^i)/(\sum_{i=1}^mn^i)=(\sum_{i=1}^mn^iB^i_lA^i_l)/(\sum_{i=1}^mn^i)$, where $W_l^i$ is the full-size LoRA of the i-th client. Clients with abundant resources can use higher LoRA ranks, thereby contributing more general, task-agnostic knowledge. This dynamic adjustment mechanism avoids the problem in traditional FL where the resources of other clients cannot be fully utilized due to the limitations of the client with the least resources. 

In addition, there are also methods based on model pruning\cite{bai2024fedspallm} and split learning\cite{tian2022fedbert,shin2023fedsplitx,xu2023federated} that can address resource heterogeneity. Bai et al. \cite{bai2024fedspallm} first proposed the pruning scheme FedSpaLLM for FLLM, which allows clients to prune the model locally based on private data while considering system resource heterogeneity and maintaining competitive communication efficiency. The server randomly samples a portion of the model layers for each client in each communication round, with the sampling quantity proportional to the client's computational resources. This proportional allocation ensures that each client can effectively participate in model training within its capabilities. Split learning, on the other hand, reduces the processing load on clients by dividing the complete model into smaller submodels. Clients can choose to run the initial layers of the model based on their computational resources, while the server is responsible for running the remaining layers. This division allows resource-constrained clients to handle only the lightweight parts of the model, leaving the computationally intensive tasks to the server.

The resource capacity of clients determines the size of the model they can train, which may lead to model heterogeneity. Samiul Alam et al. \cite{alam2022fedrolex} proposed a model-heterogeneous training scheme based on partial training called FedRolex. Through a rolling sub-model extraction scheme, it enables the parameters of the global server model to be trained uniformly, thereby mitigating the client drift problem caused by model heterogeneity.

\subsection{Data Heterogeneity}

Data heterogeneity, that is, the imbalance in data distribution across different clients, unequal data volumes, and the non-independent and identically distributed (Non-i.i.d.) nature of the data \cite{chen2024feddat}, can lead to drift in local model updates during training. This means that the update directions of models on different clients are significantly different. As a result, the global model obtained by aggregating parameters from various client models on the server is unstable and usually not the optimal model \cite{mendieta2022local,li2021align}. LLMs typically require substantial amounts of data for adequate training. However, in FL, data is distributed across multiple clients and is highly heterogeneous, making it difficult for the model to learn globally effective features from local data. Moreover, clients may only be exposed to specific patterns in their local data, leading to overfitting of local models and a decrease in the generalization ability of the global model.

Similar to resource heterogeneity, most current methods for addressing data heterogeneity are based on LoRA and Adapter. Qi et al. \cite{qi2024fdlora} proposed the FDLoRA scheme with a dual-module configuration. Each client is equipped with two LoRA modules, one for capturing local personalized knowledge and the other for global knowledge. Only the parameters of the global LoRA module are shared during parameter sharing. Clients then adaptively fuse the parameters of the dual LoRA modules based on the aggregated parameters from the server, achieving good performance, especially when the degree of data heterogeneity among clients is severe. Fang et al. \cite{fang2024automated} proposed the FedPipe scheme, whose core idea is to configure different LoRA matrices for each client. FedPipe first identifies the weights that contribute most to model training. By analyzing the importance of these weights, it selects the weights suitable for fine-tuning on edge devices. For each selected weight, FedPipe configures a LoRA that is trained on local data, dynamically adjusting the LoRA's parameters (such as batch size and decomposition rank) according to the computational and storage resource constraints of each edge device. This method ensures that each edge server can efficiently fine-tune the model within its resource limitations. Jiang et al. \cite{jiang2024personalized} proposed the PFIT method, which uses reinforcement learning to fine-tune local LLMs. Each client adjusts model parameters using different reward models based on the characteristics and needs of its local data. Clients use a global adapter to fine-tune the global model. The role of the global adapter is to enable the model to adapt to the characteristics of global tasks, but its parameter updates are fed back to the server for updating the global model. Meanwhile, clients train local LoRA on local data to achieve personalized adjustments. The parameters of these LoRAs are updated locally and are not uploaded to the server, thereby eliminating the heterogeneity of client data.

In addition, there are also some schemes that consider the training process and propose innovative methods. Panchal et al. \cite{panchal2024thinking} proposed SPRY, which uses Forward-mode Auto-Differentiation to fine-tune LLMs, achieving low memory usage, high accuracy, and fast convergence. When the data among clients are homogeneous, the global gradient aggregated on the server side by SPRY is an unbiased estimator of the true global gradient. Although heterogeneity increases the bias of the estimator, it remains usable. Mei et al. \cite{mei2024fedmoe} addressed the challenges posed by data heterogeneity by employing a Mixture-of-Experts (MoE) model. The proposed FedMoE constructs an optimal sub-MoE model for each client and feeds knowledge back into the global MoE to improve efficiency in data heterogeneity environments.

In the FLLM environment, data heterogeneity is a key challenge. However, most of the current methods for addressing data heterogeneity focus on model architecture design, with fewer methods employing other techniques such as model distillation. We believe that more universal and efficient FLLM solutions for data heterogeneity will emerge soon.

\subsection{Task Heterogeneity}

Task heterogeneity refers to the fact that different clients may have different types of tasks. In real-life FLLM scenarios, computational and storage resources are very limited. Training separate models for each task would significantly increase the computational and storage burden, potentially exceeding the resource limitations of the devices. By sharing a single model, multiple tasks can be handled simultaneously within limited resources, improving resource utilization. For example, a LLM can handle tasks such as text classification, sentiment analysis, and machine translation simultaneously, without the need to train and deploy separate models for each task. Multi-task learning can enhance the model's generalization ability because it forces the model to learn more generic feature representations instead of overfitting to the data of a specific task. This generalization ability enables the model to perform better when facing new tasks. These advantages make multi-task learning a more practical and efficient choice in resource-constrained and privacy-sensitive environments. However, multi-task learning is prone to model forgetting issues \cite{liu2022backdoor}, and conflicts in optimization objectives as well as differences in model structures pose significant challenges to the stable convergence of the global model.

Currently, research on task heterogeneity in FLLM is relatively limited, and the focus is on expanding model scale by setting up multiple adapters. Zhang et al. \cite{zhangenhancing} proposed the M2FEDSA framework, which combines split learning and multi-modal FL. It introduces a dual adaptive fine-tuning strategy by adding task adapters in the high-level encoder on the main server and modality adapters in the low-level encoder on the client side to enhance the model's adaptability to different tasks and modalities. It also employs a dual knowledge transfer strategy to pass multi-modal knowledge to single-modal features at the feature and decision levels, further improving model performance. Yang et al. \cite{yang2024dual} proposed the FedDPA framework, which combines global and local adapters to learn general knowledge across different distributions and provide personalized services for each client. Additionally, FedDPA introduces an instance-based dynamic weighting mechanism that dynamically integrates global and local adapters during inference to achieve effective test-time personalization. Chen et al. \cite{chen2024fedbone} proposed the FedBone framework, which splits the model into a general model deployed on the cloud and a task-specific model deployed on the client side. The cloud's powerful computing capabilities handle general feature extraction, while the client is only responsible for lightweight data embedding and task output. FedBone also introduces the GPAggregation method, which calculates the attention values of task gradients and historical aggregated gradients and performs projection operations to eliminate conflicts between gradients of different tasks, enhancing the model's generalization ability. It designs a task adaptation module using deformable convolution and self-attention mechanisms to further enhance the model's adaptability to different tasks. Chen et al. \cite{chen2024feddat} proposed the FedDAT scheme, which effectively addresses task heterogeneity through a dual-adapter teacher module and mutual knowledge distillation strategy, achieving distributed fine-tuning of the base model while maintaining communication efficiency. Experimental results show that FedDAT significantly outperforms existing centralized PEFT methods in multiple multi-modal FL benchmarks, demonstrating better convergence speed and scalability. Ping et al. \cite{ping2024fl} proposed a FLLM training method called FL-TAC, which trains a low-level adapter for each individual task on the client side and then clusters similar adapter groups on the server side for task-specific aggregation. 

The heterogeneity issues of FLLM are primarily manifested in resource heterogeneity, data heterogeneity, and task heterogeneity. The ability to address these heterogeneity issues in FLLM is crucial for their robustness and has become one of the current research hotspots. Presently, most research focuses on the resource heterogeneity of FLLM, while further in-depth studies are needed for data and task heterogeneity.

\section{Security of FLLM}
The primary motivation for combining LLMs with FL is to enable LLM training that collects data from various parties while protecting user privacy. Therefore, privacy protection is the most important feature of FLLM. However, due to the emerging nature of FLLM, current research has focused excessively on how to achieve efficient fine-tuning of FLLM, with insufficient analysis of the system's privacy and security. Both FL and LLM have their own privacy and security threats. Whether the combination of the two will lead to an accumulation of all risks or mitigate some attacks has not yet been clearly summarized in any review. This section analyzes whether the attacks that originally existed separately in FL and LLM are still effective for FLLM and surveys the new risks generated by the combination of FL and LLM. After conducting an extensive search, we found that research on attacks targeting FLLM remains limited. Correspondingly, studies on defense mechanisms for FLLM are also scarce, posing significant challenges to its security.

\begin{figure}[htbp]
  \centering
  \includegraphics[width=\linewidth]{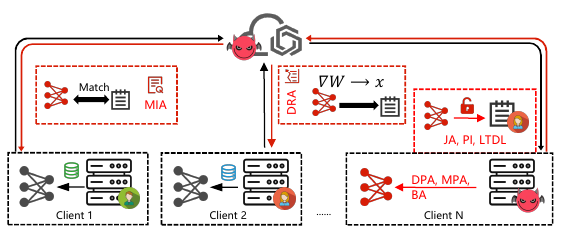}
  \caption{Privacy and security threats faced by FLLM system. Honest-but-curious server steals client privacy through MIA and DRA. Malicious clients steal other users' privacy through JA, PI, and LTDL, and also damage or manipulate the aggregation of the global model through DPA, MPA, and BA.}
  \label{fig:attack}
\end{figure}

In FL systems, although the server cannot access the raw data of clients, it may still be able to infer client privacy through the collection of model parameters or gradients. For example, membership inference attacks \cite{vu2024analysis} and input reconstruction attacks \cite{zheng2023input} are potential threats. Malicious clients, aiming to control the direction of global model training or to prevent model convergence, may launch data poisoning attacks \cite{wang2020attack,xie2019dba,zhang2022dim} and model poisoning attacks \cite{bagdasaryan2020backdoor,bhagoji2019analyzing,fung2020limitations,sun2019can}, as shown in Figure ~\ref{fig:attack}. LLMs are trained on vast amounts of public data, which may include private information such as email addresses and phone numbers. As high-quality public data becomes scarce, training LLMs on private data is becoming more common, potentially exposing private data to privacy breaches. Prominent methods for privacy theft include membership inference and reconstruction attacks. Additionally, during training, LLMs not only learn the underlying logic of language but also potentially acquire knowledge that contradicts human values, posing the risk of generating harmful content. Jailbreaking attacks are a primary means that may trigger this risk. Furthermore, the combination of FL and LLMs introduces new privacy and security threats. Training data for LLMs can inadvertently be reflected in generated content, potentially revealing sensitive personal information such as medical records and bank account numbers. This allows other clients in the FL system to potentially steal user privacy. Due to the deep transformer architecture and multi-stage training process of LLMs, poisoning attacks are more likely to succeed and are harder to detect in FLLM training. The high reliability of FLLM is a key factor for their successful implementation. We provide an overview of privacy and security attacks and defenses of FLLM in Tables ~\ref{priva} and ~\ref{secur}, respectively.

\subsection{Privacy Leakage and Defenses}

\begin{table}[htbp]
\renewcommand{\arraystretch}{1.2}
\caption{Overview of privacy in FLLM.}
\begin{center}
\addtolength{\tabcolsep}{0pt}
\begin{tabular}{c|c|c|c}
\hline

\multicolumn{2}{c|}{\textbf{Type}}& \textbf{Describe} & \textbf{Approaches} \\
\hline

\multirow{8}{*}{Attacks}& \makecell{Membership\\ Inference\\ Attack} & \makecell{Determine whether a specific data point was used.}  &\cite{vu2024analysis}\\
\cline{2-4}
& \makecell{Data\\ Reconstruction\\ Attack} & \makecell{Recover original data from model outputs or parameters.}  &\cite{petrov2024dager}, \cite{zheng2023input}\\
\cline{2-4}
& \makecell{Jailbreaking\\Attack} & \makecell{Exploit vulnerabilities in system constraints to gain\\ unauthorized access.}  &\cite{li2024peft}\\
\cline{2-4}
& \makecell{Prompt\\ Injection} & \makecell{Manipulate input prompts to elicit malicious outputs.}  &-\\
\cline{2-4}
& \makecell{Long-Tailed\\ Data Leakage} & \makecell{Infer sensitive information from underrepresented classes\\ in imbalanced datasets.}  &-\\
\hline

\multirow{7}{*}{Defenses}& \makecell{Multi-\\layered \\defense} & \makecell{Adapt multi-layered defense mechanism, such as weighted \\aggregation, geometric median, model pruning, and noise\\ addition to resist attacks.}  &FedSecurity \cite{han2024fedsecurity}\\
\cline{2-4}
& \makecell{Parameter \\quantization}  & \makecell{Integrating quantization and LoRA to exchange only partial\\ model parameters during training.}  &FedLPP \cite{zhu2024promoting}\\
\cline{2-4}
& \makecell{Differential\\ privacy}  & \makecell{Combine differential privacy with adapter mechanisms.}  &FedPA \cite{zhang2024federated}\\
\cline{2-4}
& \makecell{Local\\fine-\\tuning} & \makecell{Fine-tuning LLMs directly on client devices with private data.}  &Titanic \cite{su2024titanic}\\

\hline
\end{tabular}
\label{priva}
\end{center}
\end{table}

\subsubsection{Membership Inference Attack}

Membership Inference Attack (MIA) aims to determine whether a target sample exists in a specific participant's training set \cite{shokri2017membership}. Bai et al. \cite{bai2024membership} categorized the implementation methods of membership inference attacks in FL into update-based and trend-based approaches. One update-based method involves using model gradients as attack feature vectors \cite{nasr2019comprehensive,gupta2021membership} or comparing the differences in gradients across different rounds \cite{melis2019exploiting,li2023effective} to infer membership status. However, given that FLLM training requires datasets on the scale of billions, the presence or absence of a single training data point may have a negligible impact on gradient changes, making it difficult to determine whether it has been trained on. This has been confirmed in the research by Duan et al. \cite{duan2024membership}. They conducted a large-scale evaluation of a series of language models (ranging from 160M to 12B parameters) and found that in most cases, the performance of membership inference attacks only slightly outperformed random guessing, which is in stark contrast to previous findings in traditional machine learning models. This inefficiency is primarily attributed to two characteristics of LLMs: first, the use of massive amounts of data for training, which makes it difficult for the model to overfit the training data; and second, the training process typically involves only a little over one epoch, resulting in strong model generalization and difficulty in distinguishing between training and non-training data.

Another update-based method infers membership status through structure modifying \cite{pichler2023perfectly,nguyen2023active}. A malicious server can meticulously design the model structure, embedding malicious parameters that activate when the target member participates in training, thereby leaking membership information. Although this method places high demands on how the server designs the model structure, similar attacks have been realized in FLLM \cite{vu2024analysis}. Minh N. Vu et al. \cite{vu2024analysis} proposed two methods for a malicious server to launch active membership inference attacks in the context of FLLM. The first involves setting the first two layers of the model as fully connected layers and carefully designing the weights and bias parameters based on the target sentence or token. The presence of zeros in the output of these layers determines whether the target exists in the input. The second focuses on the self-attention layer, similarly designing the parameters of the $Q$, $K$, and $V$ matrices of the self-attention layer based on the target token and analyzing the output values to determine the target member's attributes. However, the paper assumes that the attacker can fully control the global model and manipulate its architecture, which may not entirely hold in practical applications.

Trend-based methods leverage the change trajectories of model outputs or parameters across multiple rounds to infer membership information \cite{zhang2023efficient,zari2021efficient}, requiring only the observation of metric changes in each round with minimal time overhead. This method is simple and efficient to implement and could potentially be realized in FLLM, although such instances have not yet appeared.

\subsubsection{Data Reconstruction Attack}

Data Reconstruction Attack (DRA) is a type of attack that reconstructs original data from gradients or other information. Yang et al. \cite{yang2023gradient} categorized gradient leakage attacks in FL into two major types: optimization-based attacks and analysis-based attacks, and proposed a new generation-based GLA paradigm, demonstrating its advantages in terms of data reconstruction performance and efficiency. Optimization-based attacks generate initial data randomly and optimize the data based on gradients to make it close to the private training data, thereby achieving data reconstruction. For example, Mu et al. \cite{zhu2019deep} studied a gradient-based reconstruction attack algorithm, mainly using deep learning techniques and algorithms to analyze the gradient information in FL models to recover the original training image data. They proposed an algorithm called deep leakage from gradient (DLG), which can recover the original image by generating virtual image labels and calculating virtual gradients to match the real gradients, without access to the original dataset. This optimization-based method, relying solely on gradient changes, remains applicable to the fine-tuning methods of LLM parameter updates. In contrast, analysis-based attacks solve a system of linear equations to obtain the original private data, which is more accurate but only suitable for smaller models \cite{phong2017privacy,fowl2021robbing,zhu2020r}. For LLMs with a huge number of parameters, this type of attack is more difficult to implement.

In the context of FLLM training, adversaries attempt to recover users' input text, and the discrete nature of this input increases the difficulty of data reconstruction attacks. Zheng et al. \cite{zheng2023input} explored the security issues of implementing vertical FL for LLMs in a white-box scenario, pointing out that attackers can easily and at low cost reconstruct users' input text from the intermediate embedding layers, and discussed several possible solutions to enhance the privacy protection of vertical federated LLMs. Petrov et al. \cite{petrov2024dager} proposed DAGER, a gradient inversion attack algorithm targeting FLLM, capable of accurately recovering entire input text batches from shared gradients. DAGER leverages the low-rank structure of gradients in self-attention layers and the discrete nature of token embeddings, employing exhaustive heuristic search and greedy methods to precisely recover batches for both encoder-based and decoder-based architectures. Lu et al. \cite{lu2022april} proposed APRIL, a novel attack method that analyzes the gradient leakage risks of self-attention mechanisms and demonstrates that attackers can use the shared gradient updates of models to recover private training data within the FL framework. The paper particularly observed that learnable positional embeddings are a weak link in the privacy protection of Transformer models. Fowl et al. \cite{fowl2022decepticons} proposed the DECEPTICONS attack method, which deploys malicious parameter vectors in FL to leak users' private text data, utilizing the characteristics of the Transformer architecture and token embeddings to extract token and positional embeddings separately to recover high-fidelity text, even in the face of small batches, multiple users, and long sequences. Rashid et al. \cite{Rashid2023fltrojan} identified which training rounds included the participation of victims using the victim round identification method and proposed the maximizing data memorization method based on selective weights optimization and weights transformation learning to further enhance the model's memorization of sensitive data, significantly increasing the success rate of private data reconstruction (up to 71\%).

\subsubsection{Jailbreaking Attack}
Jailbreaking Attacks (JA) refer to bypassing or breaking through the security and censorship functions of a model to perform unauthorized operations or output non-compliant content \cite{xu2024bag,li2024peft}, which has become one of the unique and mainstream attacks against LLMs. In the FLLM scenario, the training process involves multiple participants, each of whom can be a potential target for jailbreaking attacks. Attackers may inject malicious data into the training set or craft malicious prompts to steal other users' privacy from the global model. These attacks typically involve designing clever prompts to induce the model to exceed its preset limitations while performing tasks, thereby achieving the attacker's goal. Usually, attackers launch jailbreaking attacks on LLMs through "jailbreaking prompts." Initially, "jailbreaking prompts" were mainly designed manually, which had limitations in terms of readability and fluency, and were later improved in subsequent work \cite{jiang2025decomposition}. As jailbreaking attacks continue to evolve, diffusion models have been employed for generating "jailbreaking prompts" \cite{wang2024chain}, significantly increasing the success rate of jailbreaking attacks while optimizing the fluency and diversity of the prompts. Current defense measures require security-oriented training and adversarial training at the model level to enhance the model's resistance to attacks \cite{li2024peft}.  Although various robust FL mechanisms, such as robust aggregation schemes \cite{li2023experimental}, can protect training from malicious updates, their effectiveness against emerging jailbreaking threats remains to be explored.

\subsubsection{Prompt Injection}
Prompt Injection (PI) is a technique that uses malicious instructions as part of the input prompt to manipulate the output of a language model. In the FLLM scenario, where the models are exposed to a large number of unfamiliar users, prompt injection attacks exhibit characteristics of being difficult to predict and defend against, and having a strong immediacy. It is similar to SQL injection attacks in database security, where carefully crafted inputs bypass the model's normal processing procedures to achieve unauthorized data access, execute malicious code, or produce harmful outputs. Prompt injection can be divided into two forms: direct injection and indirect injection. The former involves directly adding malicious instructions to user input, while the latter hides malicious instructions in documents that may be retrieved or ingested by the model \cite{ivuanușcua2024impact}. This type of attack primarily affects the integrity and security of applications based on LLMs, potentially leading to unauthorized data access, execution of malicious code, or generation of harmful outputs. General defense methods involve strengthening input validation and filtering at the application level to prevent untrusted user input from being directly passed to the LLM.

\subsubsection{Long-Tailed Data Leakage}
Long-tailed data leakage (LTDL) refers to the over-memorization of rare data by a minority of participants. This characteristic is particularly dangerous in FLLM because the local data of participants often contains sensitive information (such as medical records and financial transactions), and the global model may inadvertently leak these details through parameter aggregation. On the one hand, due to the large number of parameters and multi-layer attention mechanisms in LLMs, they possess extremely strong data representation capabilities, but also face the risk of over-memorizing long-tailed data \cite{eriksen2024use}. Studies have shown that even when facing extremely few samples (such as rare case data from a certain participant), the model may accurately memorize data details through subtle changes in gradient updates. For example, Liu et al. \cite{2023Adversarial} verified a variety of adversarial attack techniques targeting LLMs and found that high attack success rates can still be achieved with a small number of attack samples, especially on LLaMA-7B, where the ASR of 8-shot attacks always remains above 50\%. 

On the other hand, the training method of FLLM itself also exacerbates the severity of this problem. Taking a trillion-parameter model as an example, a single gradient update requires the transmission of hundreds of GB of parameters, and the frequent global aggregation in FL leads to exponential growth in communication bandwidth and computational resource consumption \cite{yao2024survey}. For instance, training a model of the scale of GPT-3 requires thousands of GPUs to work in parallel for several months, and the dispersed participants in FL may be forced to reduce the frequency of aggregation due to hardware heterogeneity (such as insufficient computing power of edge devices), which exacerbates the overfitting of local models to long-tailed data. Ma et al. \cite{ma2022state} pointed out that the Non-i.i.d. data distribution will intensify the training directionality problem of the model, and the low-frequency update strategy is difficult to eliminate the parameter bias in the local model during global aggregation.

Research on the long-tailed privacy leakage risks in FLLM is currently still in its infancy. However, some foundational work on FL or LLMs can offer valuable insights. For instance, adding noise (such as Gaussian or Laplacian noise) to the gradients or parameters of FLLM can mitigate the risk of data memorization. Batool et al. \cite{batool2024secure} proposed a VANETs FL framework that implements a lightweight privacy budget allocation strategy through differential privacy design, optimizing model aggregation efficiency while ensuring privacy security. However, this solution experiences a more significant accuracy loss in Non-i.i.d. data scenarios. Additionally, Mao et al. \cite{mao2025survey} suggested reducing the number of trainable parameters through LoRA, splitting LLMs into shared and private layers, and aggregating only the shared parameters. This approach can reduce parameter scale and memory effects to some extent but still faces communication efficiency issues in models with hundreds of billions of parameters.

\subsubsection{Defenses}

The privacy protection of FLLM faces numerous challenges, including data leakage risks and model protection. Recently, researchers have proposed a variety of privacy protection schemes, ranging from quantization techniques and differential privacy to distributed training paradigms, aiming to balance the relationship between privacy protection and model performance.

In the FedSecurity framework proposed by Han et al. \cite{han2024fedsecurity}, FedDefender, as a key component, is specifically designed for the defense mechanisms of FLLM to counter various attacks. FedDefender implements defensive measures at different stages of FL training, including the "pre-aggregation," "during-aggregation," and "post-aggregation" phases. Before aggregating client models, FedDefender can score local models to identify potentially malicious ones and reweight them to mitigate the impact of malicious models. For example, the Krum algorithm tolerates a certain number of Byzantine clients by selecting the single most likely benign model as the global model. During the aggregation process, FedDefender modifies the aggregation function to make it more robust against potential malicious client models. For instance, Robust Federated Aggregation calculates the geometric median of client models as the aggregated model instead of simply averaging them. After aggregation, FedDefender can directly modify the global model by clipping or adding noise to protect it from potential attackers. For example, Clipping-based Robust FL clips the global model after each aggregation to limit the model's norm. Through these multi-layered defense mechanisms, FedDefender can flexibly respond to different types of attacks, including data poisoning, model poisoning, and data reconstruction attacks. 

Zhu et al. \cite{zhu2024promoting} proposed the FedLPP framework, which combines quantization techniques and LoRA to protect both data and model privacy in FL. FedLPP distributes quantized rather than complete model parameters during training, preventing clients from obtaining the full model on the server and effectively protecting the privacy of the global model. By updating only a small portion of the model's parameters, FedLPP further reduces communication overhead and limits clients' access to the global model's details. 

Zhang et al. \cite{zhang2024federated} suggested that LLM privacy protection can be achieved through FL frameworks and personalized adapter mechanisms. Each client learns a lightweight personalized adapter using its private data, which collaborates with the pre-trained base model to provide efficient and fine-grained services for recommendation systems. Throughout the process, users' private data remains on local devices and is not shared with the server, ensuring data privacy. The method further enhances privacy protection with differential privacy techniques, such as adding noise to model parameters when clients upload them to prevent the server from inferring users' original data. This data-localized privacy protection mechanism not only safeguards users' privacy but also allows models to integrate shared knowledge without sharing sensitive information while retaining each user's personalized preferences. 

Su et al. \cite{su2024titanic} proposed the Titanic scheme, which deploys the fine-tuning process of LLMs directly on client devices holding private data. This approach ensures that private data always remains on local devices and is not sent to the cloud or other centralized servers, thereby maximizing data privacy protection. However, this method is impractical for resource-constrained clients. To address the challenge of client resource limitations, Titanic implements fine-tuning of LLMs on client devices in four ways: (\romannumeral1) Model partitioning and distributed fine-tuning: Titanic splits the LLM across multiple client devices for fine-tuning instead of requiring each client to train the entire model independently. This significantly reduces the computational burden on individual clients, allowing resource-constrained devices to participate in model training. (\romannumeral2) Optimized client selection: Titanic first selects a subset of clients using an efficient integer optimization algorithm. These clients are more representative in terms of computational resources and data quality. In this way, Titanic ensures that the participating clients can efficiently complete the tasks assigned to them while reducing over-reliance on individual client resources. (\romannumeral3) Reduced communication overhead: Titanic significantly reduces communication costs by transmitting only a small number of model weights between clients instead of entire model updates. This not only protects privacy but also lowers bandwidth requirements, making distributed training more feasible. (\romannumeral4) Model-agnostic partitioning mechanism: Focusing on feasibility, Titanic adopts a model-agnostic partitioning mechanism that can fully automate the splitting and distribution of any LLM to client devices. This means that Titanic can flexibly adapt to different models and trainers without modifying the model source code. Through these technical means, Titanic effectively addresses the problem of client resource limitations while protecting data privacy, making it possible to fine-tune LLMs on resource-constrained devices.

The privacy protection solutions for FLLM are still in their infancy, and the specific problems they target vary. To enhance the security of FLLM, there is an urgent need for deeper and broader exploration in this area.

\subsection{Security threats and defenses}

\begin{table}[htbp]
\renewcommand{\arraystretch}{1.2}
\caption{Overview of security in FLLM.}
\begin{center}
\addtolength{\tabcolsep}{0pt}
\begin{tabular}{c|c|c|c}
\hline

\multicolumn{2}{c|}{\textbf{Type}}& \textbf{Describe} & \textbf{Approaches} \\
\hline

\multirow{5}{*}{Attacks}& \makecell{Data\\ Poisoning\\ Attack} & \makecell{Maliciously alter training data to degrade model performance.}  &-\\
\cline{2-4}
& \makecell{Model\\ Poisoning\\ Attack} & \makecell{Maliciously modify model parameters or updates to degrade the\\ overall model performance.}  &-\\
\cline{2-4}
& \makecell{Backdoor\\ Attack}  & \makecell{Insert hidden triggers into the LLM to produce incorrect outputs\\ when activated.}  &\makecell{\cite{li2023backdoor}, \cite{lyu2024lurking}, \\\cite{wu2024vulnerabilities}, \cite{li2024unveiling}}\\

\hline

\multirow{5}{*}{Defenses}& \makecell{Distance\\ based\\ defense} & \makecell{Adversaries are identified by the distance deviations in malicious\\ updates from normal ones.}  &\makecell{\cite{zhou2024secfft}, \cite{ren2024bpfl}, \\\cite{zhang2022dim}}\\
\cline{2-4}
& \makecell{Feature\\ based\\ defense}  & \makecell{Maliciously tampered model updates exhibit distinct characteristics\\ from benign updates in certain features.}  &\makecell{\cite{ma2024federated}, \cite{basak2025dpad}, \\ \cite{zhang2023fed}, \cite{ali2024adversarially}}\\
\cline{2-4}
& \makecell{ Knowledge \\Distillation}  & \makecell{Integrate clustering, model selection, and knowledge distillation\\ to identify and filter malicious client updates.}  &\cite{alharbi2025robust}\\

\hline
\end{tabular}
\label{secur}
\end{center}
\end{table}

\subsubsection{Data Poisoning Attack}

Data Poisoning Attacks (DPA) occur during the data collection phase on the client side, where the original data is modified to train a poisoned local model, which is then uploaded to participate in aggregation to harm the global model and compromise its availability or integrity. Shafahi et al. \cite{shafahi2018poison} explored an optimization-based "clean-label" data poisoning attack on neural networks, which manipulates the model's behavior at test time by adding carefully designed samples to the training set. This type of attack does not require the attacker to control the labels of the training data but instead leverages the model's "memory" of the data during training to achieve its goals. Another form of DPA is the label-flipping attack, where the adversary modifies the labels of the dataset rather than the sample features to generate a poisoned model. Tolpegin et al. \cite{tolpegin2020data} studied label-flipping attacks on FL systems, demonstrating that even a small number of malicious participants can significantly reduce classification accuracy and recall, and that the attack can be targeted to negatively impact specific categories. LLMs, with their higher complexity and stronger fitting capabilities, can capture subtle features in the data, including maliciously altered label information. Therefore, label-flipping attacks may have a more pronounced impact on LLMs, as they are more prone to overfitting incorrect label information. Shejwalkar et al. \cite{shejwalkar2022back} systematically analyzed various possible threat models, variants of poisoning attacks, and different capabilities of attackers, with a particular focus on non-targeted poisoning attacks. They found that, contrary to common belief, FL shows high robustness in practical applications even with simple and low-cost defense measures. Based on this, they proposed new state-of-the-art data poisoning attack methods and demonstrated their ineffectiveness in the presence of simple defense mechanisms through extensive experiments on three benchmark datasets. In addition to directly modifying the original data, Zhang et al. \cite{zhang2020poisongan} proposed PoisonGAN, a generative poisoning attack model for FL systems. This method, based on generative adversarial networks (GANs), uses the parameters of the global model to generate toxic data samples that mimic the training samples of other participants and forge the labels of these samples. Since the federated fine-tuning process typically involves making a small number of updates to the pre-trained model, these toxic data samples can significantly impact the model's performance during the fine-tuning stage.

\subsubsection{Model Poisoning Attack}

Unlike DPA, Model Poisoning Attacks (MPA) occur during the training phase on the client side, where the local model is modified to achieve the goal of corrupting the global training. Fang et al. \cite{fang2020local} first systematically studied MPA and formalized the attack problem as an optimization problem, targeting four byzantine fault-tolerant FL defense methods. This optimization approach helps minimize the difference between the current poisoned model and the model from the previous round, making it more difficult for the server to detect the attack. Bagdasaryan et al. \cite{bagdasaryan2020backdoor} used a model replacement method to blend the poisoned model with a benign model and employed hyperparameter scaling to evade detection. Shejwalkar et al. \cite{shejwalkar2021manipulating} proposed a more effective model poisoning attack, similar to the Min-Max attack, which constrains the upper bound of the sum of squared distances between the malicious gradient and all benign gradients to be the sum of squared distances between any benign gradient and other benign gradients, thereby ensuring the survival rate of the malicious model. Federated fine-tuning typically involves making a small number of updates to the parameters of the pre-trained model, based on the local data of the clients. Attackers can tamper with these updates to directly affect the fine-tuning process of the global model. However, MPA targeting FLLM have not yet emerged.

\subsubsection{Backdoor Attack}

Poisoning attacks degrade the performance of the global model by tampering with data and models, while backdoor attacks (BA) manipulate model behavior by injecting specific attack information or data. Li et al. \cite{li2023backdoor} investigated the threat of backdoor attacks when fine-tuning base models in FL, proposing a method to embed backdoors into the base model and transfer them into the FL system. This allows the successful implantation of backdoors in the global model without fully participating in the FL process. Yang et al. \cite{yang2021careful} explored vulnerabilities to backdoor attacks in the word embedding layer of natural language processing models. They found that attackers could inject backdoors by modifying only one word embedding vector (i.e., the embedding vector of the trigger word) without accessing the target dataset. This enables the model to produce incorrect classifications for input samples containing the specific trigger word without affecting its performance on normal samples. However, the trigger word needs to be rare and not appear in the clean test set, which may limit the practical application scenarios of the attack, as attackers need to carefully select trigger words to avoid detection. Yoo et al. \cite{yoo2022backdoor} studied the feasibility of backdoor attacks through rare word embeddings and gradient ensembling. Attackers can inject backdoors by manipulating the embedding vectors of rare words, causing the model to produce incorrect outputs for inputs containing specific trigger words without affecting its performance on normal samples. Lyu et al. \cite{lyu2024lurking} proposed a novel backdoor attack method called PFedBA, which optimizes the trigger generation process to align the gradients and losses of the backdoor task with the main task, embedding undetectable backdoors in personalized models. Wu et al. \cite{wu2024vulnerabilities} introduced a novel attack strategy that generates synthetic data on the server side using a tampered base model and implants backdoors during client model initialization and knowledge distillation. This attack method has a high success rate in various image and text classification tasks, and existing FL defense strategies have limited effectiveness against this novel attack. Xi et al. \cite{li2024unveiling} proposed a new backdoor attack called Fed-EBD, which generates backdoored synthetic data on the server side using a tampered base model and propagates it to client models without requiring the attacker to fully control clients or continuously participate in the FL process. Experiments showed that this attack has a high success rate in various heterogeneous FL configurations and benchmark datasets and can effectively evade existing backdoor defense strategies. The study revealed significant security risks when using federated models in horizontal FL and emphasized the urgency of developing more robust defense mechanisms.

However, current backdoor attacks generally have strong attack assumptions and limited experimental scopes, with insufficient consideration of practical application scenarios. They do not fully account for the potential impact of other security mechanisms (such as client authentication and data encryption) that may exist in real systems on the attacks.

\subsubsection{Defenses}
To achieve robust evaluation of model parameters from client models, researchers have conducted more in-depth explorations. Zhou et al. \cite{zhou2024secfft} proposed SecFFT, which utilizes frequency-domain transformations to extract the low-frequency components of model updates and identifies malicious updates inconsistent with the normal update distribution using chi-square distance. It also analyzes the historical behavior sequences of nodes to construct attack intentions and employs the local outlier factor algorithm to identify malicious intentions hidden behind seemingly normal behaviors. By combining these two methods, SecFFT can effectively detect complex and covert backdoor attacks while maintaining high performance and robustness in federated fine-tuning. Ma et al. \cite{ma2024federated} proposed a classifier based on persistent homology and persistent graphs, which identifies malicious clients by analyzing the topological features of neural network models. This method can efficiently detect various types of backdoor attacks even under highly imbalanced non-i.i.d. data conditions. Basak et al. \cite{basak2025dpad} proposed the DPAD scheme, which uses an auditing mechanism to check the integrity and consistency of client updates, identifying potentially maliciously tampered data or abnormal behaviors to prevent these harmful updates from affecting the global model. Ren et al. \cite{ren2024bpfl} proposed the BPFL method, which detects malicious behavior by calculating the cosine similarity between client local gradients and global gradients. Malicious gradients typically have lower similarity to the global gradient, so this similarity calculation can identify tampered gradients and prevent them from causing damage to the global model.

To enhance the robustness of FLLM systems against poisoning and backdoor attacks, researchers have successively proposed targeted defense schemes. Zhang et al. \cite{zhang2023fed} proposed the Fed-FA backdoor attack defense algorithm, which uses the f-divergence metric to estimate the differences in client data and addresses the issue of client data invisibility through a Hessian redistribution mechanism in the synthetic dataset and embedding layer. It demonstrates how to detect and exclude suspicious clients by modeling the differences in client data distributions, thereby effectively defending against backdoor attacks. However, the complexity of calculating the Hessian matrix and the f-divergence metric is high, and the defense capability for non-i.i.d. data is limited. Ali et al. \cite{ali2024adversarially} proposed a new defense mechanism called AGSD, which detects malicious client model updates by identifying adversarial biases and overly confident predictions in the attacked model. It combines clustering algorithms and client trust history to select the most trustworthy client updates for model aggregation. AGSD can effectively defend against attacks even with a very small retained dataset ($\leq$ 0.1\% training data) or when using out-of-distribution data, with minimal impact on the accuracy of clean data. However, the defense effectiveness of AGSD against certain specific types of adaptive attacks (such as low-confidence backdoor attacks) still needs further verification. Alharbi et al. \cite{alharbi2025robust} proposed the RKD (Robust Knowledge Distillation) defense mechanism, which identifies and filters out malicious client updates by combining clustering, model selection, and knowledge distillation techniques. This approach constructs a reliable model ensemble and distills the knowledge of these models into the global model. However, it requires additional computational resources, especially in large-scale FL environments. Zhang et al. \cite{zhang2022dim} proposed Dim-Krum, noting that backdoor attacks in NLP are more difficult to defend against compared to the computer vision (CV) domain. This is because NLP attacks typically have lower relative backdoor strengths, leading to poor performance of existing robust federated aggregation methods in NLP tasks. To address this, the authors proposed an improved algorithm Dim-Krum based on the Krum framework, which calculates distances between clients only in a few dimensions. This effectively detects and discards malicious updates, significantly reducing the success rate of backdoor attacks while maintaining high accuracy on clean data.

\section{Future Directions}
As FLLM continue to improve in terms of feasibility, robustness, and security, user demands are also increasing. For example, joint training under few-shot conditions, the need for unlearning techniques when users withdraw their private data or exit the system, and the protection of model IP rights by the server. These issues pose challenges to the sustainable development of FLLM.% We summarize the future directions of FLLM in Table ~\ref{Sustainability}.

\begin{comment}
\begin{table}[htbp]
\renewcommand{\arraystretch}{1.2}
\caption{Overview of Future Directions in FLLM.}
\begin{center}
\addtolength{\tabcolsep}{0pt}
\begin{tabular}{c|c|c}
\hline
\textbf{Type}& \textbf{Describe} & \textbf{Approaches} \\
\hline

Few-shot Learning & \makecell{Each client can only provide a small number\\ of data samples, making it difficult for\\ the model to learn and generalize sufficiently.}  &\makecell{FedIT-U2S \cite{ye2024leveraging}, LP-FL \cite{jiang2024low}}\\
\hline
Unlearning&\makecell{Remove the impact of specific user\\ data from the global model while minimizing \\the damage to model performance.}  &\makecell{-}\\
\hline
IP Protection&\makecell{Prevent illegal copying, tampering, or misuse\\ of the model, and ensure fair allocation.} &\makecell{-} \\
\hline

\end{tabular}
\label{Sustainability}
\end{center}
\end{table}
\end{comment}

\subsection{Few-shot learning in FLLM}

Existing FL methods, when dealing with LLMs, often require a substantial amount of labeled data to achieve effective fine-tuning, which is neither economical nor feasible in practical applications. Particularly in fields such as healthcare and finance, where data privacy and security are of utmost importance, data is often difficult to label and share on a large scale. Few-shot learning (FSL) techniques have made significant progress in reducing the demand for labeled data, but integrating them with FL to adapt to LLMs still faces many challenges. On the one hand, FSL relies on the model's ability to learn from context, and the distributed nature of FL may lead to insufficient knowledge transfer between clients, thereby affecting the effectiveness of FSL. On the other hand, how to efficiently generate and utilize a small amount of labeled data within the federated framework to enhance model performance remains an urgent issue to be resolved. Therefore, how to fully leverage the potential of decentralized data while protecting data privacy and reducing dependence on large-scale labeled data has become a key issue in current research on FLLM.

\begin{figure}[htbp]
  \centering
  \includegraphics[width=0.8\linewidth]{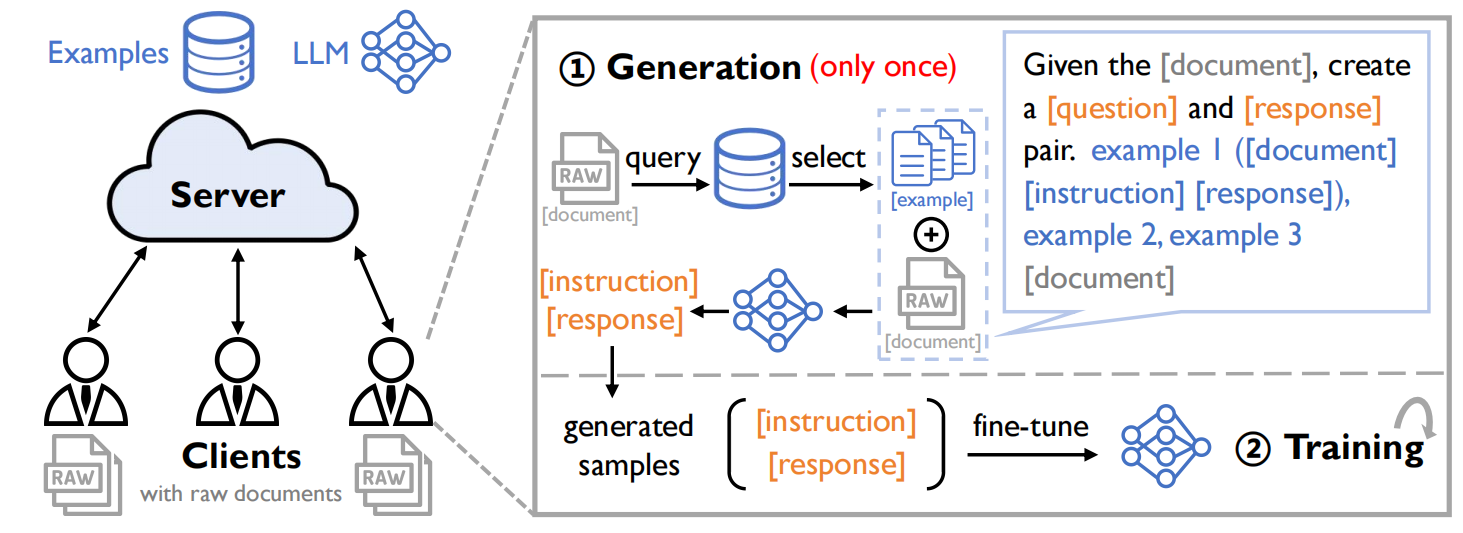}
  \caption{FedIT-U2S workflow from \cite{ye2024leveraging}. FedIT-U2S employs a retrieval-based example selection technique to automatically select the most relevant examples for each client's unstructured data from an example pool provided by the server. These examples are then combined with the client's unstructured data pieces to form prompts, which are fed into a pre-trained language model to generate instruction-response pairs. This process transforms the unstructured data into structured instruction-tuning data.}
  \label{few}
\end{figure}

Existing instruction fine-tuning methods typically assume that clients already possess structured instruction-response pair data, which is unrealistic in practice because client data is usually unstructured text. Therefore, manually annotating this data is not only time-consuming but also limits the widespread application of federated instruction fine-tuning. Ye et al. \cite{ye2024leveraging} proposed the FedIT-U2S framework, which leverages few-shot prompting techniques to combine unstructured text and a small number of examples to automatically generate structured instruction-response pair data, as shown in Figure ~\ref{few}. It also introduces a retrieval-based example selection technique that automatically selects examples based on the relevance between client data and the example pool, avoiding the complexity of manually selecting examples. Although FedIT-U2S reduces the need for manually annotated data and expands the application scope of federated instruction fine-tuning, the quality of the generated data lacks an excellent evaluation mechanism for screening, and its performance is highly dependent on the base model. Cai et al. \cite{cai2023towards} proposed the AUG-FedPrompt scheme, which designs a comprehensive data generator to simulate the distribution of labeled data in few-shot learning tasks and combines prompt learning and pseudo-labeling techniques to augment data using a large amount of unlabeled data. However, while this method achieves high performance, it also brings significant system overhead, including high computational latency, large memory requirements, and high communication costs, and its robustness to LLMs is debatable. Building on this, to address the issue of scarce and unevenly distributed data labels, they proposed the FeS framework \cite{cai2023federated}. Unlike \cite{cai2023towards}, FeS focuses more on system-level optimization, significantly reducing training latency, device energy consumption, and network traffic through curriculum training progress control, representative diversity filtering, and co-planning of training depth and layer capacity, while maintaining model performance comparable to full dataset fine-tuning \cite{cai2023federated}. Jiang et al. \cite{jiang2024low} proposed the LP-FL framework, which guides LLMs to understand task objectives by adding task descriptions to input samples. This allows clients to leverage the global model's knowledge to assign soft labels to unlabeled data and gradually incorporate it into the training set, thereby dynamically expanding the labeled dataset during the FL process.

\subsection{FLLM Unlearning}
The goal of machine learning is to extract knowledge from data, while machine unlearning endows models with the ability to "forget" specific data. Its core lies in adjusting model parameters to achieve the effect of certain data not participating in training, thereby avoiding retraining \cite{ma2022learn}. This approach allows the server to remove the contributions of specific user data according to user requests, ensuring that the model cannot trace these data, thereby protecting privacy. Meanwhile, model updates resulting from forgetting erroneous or low-quality data can further enhance model security.

In FL, simply removing the updates of the target user from the global model is not sufficient, as other users' historical models still retain the data to be forgotten, and these data will be re-aggregated in subsequent training. Large-scale data forgetting may lead to catastrophic forgetting \cite{liu2022backdoor} significantly reducing model performance. Zhang et al. \cite{zhang2025model} proposed CGKD, which is specifically designed to address the model recovery issue in federated unlearning, particularly under scenarios with limited server data resources. CGKD constructs the unlearning model by erasing all historical contributions of the target client and treats it as the student model. It then fine-tunes the pre-trained CLIP model using a small number of samples on the server side to generate a more robust teacher model. During the fine-tuning process, the original backbone network of CLIP is kept intact, and an adapter module is introduced to dynamically integrate the fine-tuned features with the original features through residual connections, thereby enhancing the model's understanding of the semantic context of images. This approach effectively mitigates the negative impact of unlearning operations on model performance. Zuo et al. \cite{TrustChain} proposed a blockchain-based federated learning framework for LLMs, which leverages the tamper-proof and distributed ledger features of blockchain to create an immutable record of each model’s contributions, thereby enhancing transparency and accountability. This function is seamlessly integrated with the federated learning mechanism, allowing data owners to remove their data from the training process while minimizing the impact on other participants. The mechanism is implemented through blockchain’s smart contracts, ensuring the security and transparency of the unlearning process. Liu et al. \cite{liu2022right} proposed an efficient federated unlearning method called Rapid Retraining, which uses fast retraining and a distributed Newton-type update algorithm. It leverages the diagonal empirical Fisher information matrix to approximate the inverse Hessian vector and introduces momentum techniques to achieve data deletion while reducing errors and enhancing model utility. This method is model agnostic and can be combined with the optimization techniques commonly used in federated fine-tuning (such as LoRA), and it is possible to implement it in LLMs. Su et al. \cite{su2023asynchronous} proposed a novel asynchronous federated unlearning mechanism called KNOT, which divides clients into multiple clusters and performs aggregation only within each cluster. This approach confines the retraining caused by data deletion to within the cluster. To optimize the assignment of clients to clusters, the authors formulated the problem as a solvable optimization problem, namely the lexicographic minimization problem, and demonstrated that it can be efficiently solved using a linear programming solver, significantly reducing time overhead. This asynchronous federated learning method can improve efficiency in FLLM. However, although clustering reduces the number of clients that need to be retrained, the cost may still be high in the FLLM scenario. Zhu et al. \cite{zhu2023heterogeneous} proposed the FedLU framework for learning and offloading heterogeneous knowledge graph embeddings. Based on cognitive neuroscience theory, they proposed an offloading method that combines retroactive interference and passive decay. This method can delete specific knowledge from both local clients and the global model without significantly affecting overall performance, meeting the needs of privacy protection and data deletion. This method is promising for FLLM. Clients can utilize their local data to fine-tune pre-trained LLMs and then transfer the local knowledge to the global model through knowledge distillation.

For complex LLMs, fine-tuning is prone to catastrophic forgetting, which is the loss of old knowledge when learning new tasks \cite{zhai2023investigating}. This limits the generality and scalability of multi-task learning. Zhu et al. \cite{zhu2024model} introduced a post-training adjustment method called "Model Tailor." This method retains the pre-trained parameters of LLMs while replacing a small portion ($\leq$ 10\%) of the fine-tuning parameters. Model Tailor employs a second-order analysis-based approach to evaluate the importance of each parameter and selectively modifies those parameters that have the least impact on both the target and original tasks, thereby ensuring that the model retains most of the pre-trained knowledge after fine-tuning. Li et al. \cite{li2024revisiting} revealed a direct link between the flatness of the model loss landscape and the degree of catastrophic forgetting. Based on this connection, they introduced the Sharpness-Aware Minimization method to flatten the optimization landscape, attempting to maintain the model's memory of previous knowledge during fine-tuning, thereby alleviating the model forgetting issue. Lee et al. \cite{lee2024bapo} proposed a new method called Base-Anchored Preference Optimization. The core idea of BAPO is to maintain the possibility of the policy model generating base responses originating from the reference model during the process of personalized preference optimization. By introducing an anchoring mechanism for base responses in the optimization process, BAPO can ensure that the policy model does not lose the knowledge contained in the base responses when adapting to different user preferences, thereby effectively alleviating the problem of knowledge forgetting. These unlearning methods for LLMs may be inefficient in FLLM, as when a client initiates a request to forget data, the server cannot simply unlearn on the global model alone but also needs to ensure that the historical models of other clients forget this data as well. This can likely only be achieved through multiple iterations slowly, posing a significant challenge to the efficiency of unlearning.

Although there has been much research on machine unlearning in FL and LLMs, there is still a gap in the emerging issue of data forgetting in FLLM. Current federated unlearning methods are not only computationally and communicatively expensive for LLMs but also difficult to transplant to LoRA and adapter architectures. How to implement federated unlearning schemes for LLMs while avoiding catastrophic forgetting is of high research value in the coming period.

\subsection{IP Protection in FLLM}

Given the high training costs of LLMs, managing the authorized use of models becomes particularly crucial, and cost-effective model watermarking techniques can protect model intellectual property (IP) rights and prevent models from being illegally copied or misused. Liu et al. \cite{liu2024survey} summarized text watermarking techniques for LLMs and found that the advanced semantic understanding and context-aware capabilities of LLMs make watermark embedding more covert while reducing the impact on the original text semantics. Embedding watermarks in LLM-generated text can effectively track and detect LLM-generated text, helping to control potential misuse.

Yang et al. \cite{li2022fedipr} proposed the FedIPR framework, which allows users to independently embed private watermarks in their local models and verify these watermarks after model aggregation to prove IP rights over the federated model. FedIPR implements feature-based watermarking, embedding binary strings in the parameters of the model's normalization layers as watermarks, and backdoor-based watermarking, introducing specific trigger samples (such as adversarial samples) during model training so that the model outputs specific incorrect labels when receiving these trigger samples, thereby verifying ownership.

LLMs typically have complex structures and a large number of parameters, providing more space for watermark embedding. For example, the normalization layers of the transformer architecture can be used for feature-based watermark embedding, while adversarial samples can serve as triggers for backdoor-based watermarking. Liao et al. \cite{liao2024dynamic} noted that in heterogeneous FL, watermarks embedded in the global model may be damaged to varying degrees when transferred to users' heterogeneous models, failing to provide complete ownership protection in local models. Therefore, they proposed the PWFed method to protect model IP rights in heterogeneous FL. PWFed uses GAN technology to generate dynamic watermark samples that are indistinguishable from original samples and designs two different granularity watermark embedding strategies to ensure the robustness and stealth of watermarks in personalized models. However, in the context of LLMs, PWFed may require greater computational overhead, and its robustness remains to be considered.

So far, there has been a considerable amount of literature on IP protection for FL and LLMs, but there is still much room for exploration regarding the emerging FLLM. In the FL environment, how to ensure the legal and authorized use of LLMs and prevent unauthorized copying and dissemination is an urgent issue to be addressed.

\section{Conclusion}

With the increasing popularity of LLMs among the general public, the demand for training data has surged exponentially, while public data resources are gradually being depleted. Directly using users' private data for training would severely violate privacy. Against this backdrop, the integration of LLMs and FL has emerged and is gaining increasing attention, with the potential for broader applications in the future. However, the field of FLLM is still in its infancy, with key issues that need to be addressed urgently. Starting from the temporal overhead, heterogeneity, security and privacy issues, and other special issues of FLLM, we have discussed the cutting-edge research on the feasibility, robustness, security, and future directions of current FLLM and found the following characteristics:

1. Research on the feasibility of FLLM has become increasingly sophisticated academically but still has a significant gap from practical application. The computational overhead of client training and the communication overhead required for transmitting models during federated fine-tuning of LLMs can be reduced by hundreds or even thousands of times compared to full-parameter tuning through PEFT and some special methods. However, due to the large base of these overheads, even with such reductions, it remains challenging for ordinary participants to train LLMs on local devices. In the future, with the continuous optimization of computing resources and the development of distributed training technologies, the fine-tuning efficiency of FLLM is expected to improve further, especially with efficient training and optimization in large-scale distributed environments becoming a research focus.

2. Research on the robustness of FLLM is being actively conducted but still needs to address challenges from multiple aspects. The current research mainly focuses on resource heterogeneity, due to the high resource threshold for training LLMs and the significant differences in training capabilities among users, which easily leads to resource heterogeneity. For data and task heterogeneity issues, there are currently few solutions, and it is evident that most of them are based on LoRA and Adapter. In the future, the robustness of FLLM will gradually become a research focus after feasibility, and solving various realistic heterogeneity issues will be a key step for the practical application of FLLM.

3. Research on the security of FLLM has just begun and urgently needs to focus on threats and defenses related to privacy and security. Current research on the security of FLLM mainly focuses on backdoor attacks, while other potential threats such as data reconstruction attacks, jailbreaking attacks, and poisoning attacks are still very rare. We have found that due to different fine-tuning methods in FLLM training, the related attack and defense schemes also vary. Currently, there is limited research on various attack and defense schemes targeting different fine-tuning methods of FLLM. It can be anticipated that the field of privacy and security for FLLM will exhibit a wide range of involvement and diverse solutions.

4. Future directions of FLLM holds significant potential. For instance, the few-shot learning problem is particularly salient for FLLM, which necessitates a substantial amount of training data; federated unlearning of LLMs serves as a safeguard for users' rights to delete data; and the issue of IP protection for FLLM profoundly affects the enthusiasm of all parties involved in the training process. Research on these aspects of FLLM is currently nascent, with only a handful of papers addressing these issues. Moving forward, as these technologies continue to evolve and converge, FLLM is anticipated to better accommodate diverse complex scenarios, thereby facilitating sustainable development.

By further exploring the synergistic relationship between FL and LLMs, the field of FLLM can be advanced, leading to the development of more efficient, effective, secure, privacy-preserving, and personalized LLMs. This integration has the potential to transform artificial intelligence across various fields and promote the deployment of powerful and ethically responsible advanced AI systems.

%%
%% The next two lines define the bibliography style to be used, and
%% the bibliography file.
\bibliographystyle{ACM-Reference-Format}
\bibliography{sample-base}

%%
%% If your work has an appendix, this is the place to put it.

\end{document}